\documentclass[fleqn,usenatbib]{mnras}

\usepackage[T1]{fontenc}
\usepackage{ae,aecompl}

\usepackage{graphicx}	% Including figure files
\usepackage{amsmath}	% Advanced maths commands
\usepackage{amssymb}	% Extra maths symbols

\usepackage{color}

\title[WD1145+017 Photometric Observations]{WD 1145+017 Photometric Observations During 8 Months of High Activity}

\author[Gary et al.]{
B.L.~Gary,\thanks{Email: BLGary@umich.edu}
S.~Rappaport,\thanks{Email: sar@mit.edu}
T.G.~Kaye,\thanks{Email: tom@TomKaye.com}
R.~Alonso\thanks{Email: ras@iac.es}
and F.-J.~Hambsch \thanks{Email: hambsch@telenet.be}
\\
$^{1}$Hereford Arizona Observatory, Hereford, AZ 85615; BLGary@umich.edu \\
$^{2}$Department of Physics, and Kavli Institute for Astrophysics and Space Research, M.I.T., Cambridge, MA 02139, USA; sar@mit.edu \\
$^{3}$Raemor Vista Observatory, 7023 E. Alhambra Dr., Sierra Vista, AZ 85650; tom@TomKaye.com \\
$^{4}${Instituto de Astrof\'\i sica de Canarias, E-38205, and Dpto. de Astrof\'\i sica, Universidad de La Laguna, 38206, La Laguna, Tenerife, Spain} \\
$^{5}${Vereniging Voor Sterrenkunde (VVS), ROAD Observatory, Mol, Belgium} \\ %12 Oude Bleken, Mol, 2400, Belgium \\ 
}

\date{Accepted 2016 November 16. Received 2016 November 11; in original form 2016 July 21}

\pubyear{2016}

\begin{document}
\label{firstpage}
\pagerange{\pageref{firstpage}--\pageref{lastpage}}
\maketitle

\begin{abstract}
WD 1145+017 was observed from 2015 November to 2016 July for the purpose of characterizing transit behavior of the white dwarf by dust clouds thought to be produced by fragments of an asteroid in close orbit with the star. Fortuitously, most of these observations were carried out during a time when the overall `dip' activity was dramatically enhanced over that during its discovery with {\em Kepler} K2.  By the end of our reported observations the dip activity had declined to a level close to its K2 discovery state. Three notable events were observed. In 2016 January a large number of dust clouds appeared that had an orbital period of 4.4912 hours, and this event also marked the end of a 3-month interval dominated by the K2 `A' period. The second event was a 2016 April 21 appearance of four dip features with drift lines in a waterfall (date vs. phase) diagram that diverged from their origin date, at a location away from the `A' asteroid, and which lasted for two weeks. The third event was the sudden appearance of a dip feature with a period of 4.6064 hours, which is essentially the same as the K2 `B' period. The evolution of dip shape, depth, and total fade amount provide constraints on dust production and loss mechanisms. Collisions can account for the sudden appearance of dust clouds, and the sudden increase in dust amount, but another mechanism for continual dust production is also required. 
\end{abstract}

% Select between one and six entries from the list of approved keywords. Could we add ``debris disk''?

%%%%%%%%%%%%%%%%%%%%%%%%%%%%%%%%%%%%%%%%%%%%%%%%%%

%%%%%%%%%%%%%%%%% BODY OF PAPER %%%%%%%%%%%%%%%%%%

\section{Introduction}

It has been known for some time now that the atmospheres of about 25-50\% of all white dwarfs are polluted with the presence of metals such as Mg, Al, Si, Ca, Ti, Cr, Mn, Fe, and Ni.  The relatively short gravitational settling times of these heavy elements indicate that refractory materials have either been recently deposited onto the surfaces of these stars, or that there is a nearly continual process of the accretion of such material (e.g., Zuckerman et al.~2010; Koester et al.~2014). The composition of the pollutants is consistent with the accretion of material from rocky objects rather than cometary bodies (Zuckerman et al.~2007; G\"ansicke et al.~2012; Farihi et al.~2013).  A smaller fraction of these polluted white dwarfs (Zuckerman et al.~1987) are also found to have infrared signatures of dusty disks orbiting them (Kilic et al.~2006; Farihi et al.~2009; Barber et al.~2012; Rocchetto et al.~2015). In addition, a growing number of white dwarfs are being found with double-peaked emission lines of the Ca II triplet (854 nm) indicating a close circumstellar gas disk (see, e.g., G\"ansicke et al.~2006; Wilson et al.~2014; Manser et al.~2016).  At the present time, it appears that the most plausible explanation for the atmospheric pollution is the accretion of the debris of rocky bodies that are the collisional or tidal breakup products of planetesimals that remain from the white dwarf progenitor (Debes \& Sigurdsson 2002; Debes et al.~2012; Mustill et al.~2014; Veras et al.~2014; 2015). The dust that is seen in some of these systems is likely a byproduct of the disintegration of the accumulated rocky bodies. 

A significant boost to our understanding of this scenario was made by Vanderburg et al.~(2015) who reported the discovery of periodic transits of the white dwarf WD 1145+017 (hereafter, WD1145) using data from the K2 mission (Howell et al.~2014).  Vanderburg et al.~(2015) detected six distinct periodicities in the K2 data ranging from 4.5 to 4.9 hours (designated as ``A'' through ``F'').  The folded lightcurves corresponding to these periodicities typically exhibited 1-2 hour long dips with depths ranging up to 2\% of the mean flux level.  The determination of the transit shapes and depths were limited by the 1/2-hour long-cadence exposure times of K2 and the need to superpose many dips in order to produce a statistically significant signal. The six periodic signals appeared to be coherent over the $\sim$80 days of the K2 observations.

Since the K2 discovery there have been a number of follow-up ground-based observations of WD1145 (Vanderburg et al.~2015; Croll et al.~2015; G{\"a}nsicke et al.~2016; Rappaport et al.~2016; hereafter R16). The earlier of these observations (during 2015 April and May) found relatively infrequent transits, but with greater depths (up to 40\%) and more narrow widths of $\sim$5-10 minutes (Vanderburg et al.~2015; Croll et al.~2015).  The periodicity of these dips was difficult to measure accurately because of the limited coverage, but it was clear that the dips that were observed did not phase up to any of the K2 periodicities.  

The later follow-up observations of G{\"a}nsicke et al.~(2016) and Rappaport et al.~(2016) found a much higher rate of dip activity with a pattern that repeated at 4.5-hour intervals. These dips reached depths of $\sim$60\% and were sometimes so numerous that they overlapped to produce broad, i.e., hour-long, depressions in the flux. The dips during 2015 November through 2016 January could be tracked coherently over periods of weeks to months, and at least a dozen independent periods in the range of 4.490 to 4.495 hours were identified. In addition R16 also reported a period of 4.5004 hours that they identified with the K2 `A' period.  Because of the stability of these periods, it could be concluded that the dust clouds must be emanating continually from sizable bodies, i.e., with minimum masses of $10^{17}$ g to maximum masses of that of Ceres at $10^{24}$ g (see R16). If there were no such body releasing the dust, the inferred dust clouds would quickly shear out within a few dozen orbits.

In 2015 April, Xu et al.~(2016) made high-resolution spectroscopic measurements of WD1145, and discovered numerous circumstellar metal absorption lines.  The unique and impressive thing about these lines was their broad widths, all of which extended from $\approx -100$ to +200 km s$^{-1}$.

In an attempt to measure the size of the attenuating dust particles, Croll et al.~(2015), Alonso et al.~(2016), and Zhou et al.~(2016) measured the dip depths at several different wavelengths.  The conclusions were that the dust particles had to be $\gtrsim$ 0.5, 0.8, and 1 $\mu$m, respectively, for the three sets of observations.

To summarize some basic information of what is known about the WD1145 system, the host star has an effective surface temperature of 15,900 K (Vanderburg et al. 2015); the atmosphere is mostly He, with some H detected (Xu et al.~2016). The star's mass and diameter are estimated to be 0.6 $M_\odot$ and 1.34 $R_\oplus$. The distance is 174 parsec, and its V-magnitude is 17.2. The `A'-orbit planetesimal was estimated by R16 to have a radius of 200 km, so at this time it can be considered to be an asteroid. The `A' asteroid orbital period is 4.5004 hours, so its mean distance from the white dwarf is 0.0054 AU, or 94 times the radius of the WD. From the perspective of the `A' asteroid the white dwarf has an apparent diameter of 1.26 degrees. The orbital inclination angle would have to be $>89.4^\circ$ for bodies in the `A' orbit to transit the WD star. 

In this work we use two shorthand terms which are both convenient and reflect some of the ideas developed in R16.  We are neither certain that these notions have been confirmed by the current work, nor have they been ruled out.  We refer to the ``A asteroid'' as a possible few-hundred km body orbiting at the ``A period'' found during the K2 observations to be 4.4989 hours (Vanderburg et al.~2015; or indistinguishably, 4.5004 hours).  As estimated in R16, based on mass loss rates and lifetimes of weeks, fragments much smaller than the ``A'' asteroid  (e.g., $\sim$1 km) are thought to have broken off the main asteroid body from its L1 point and are orbiting with $\sim$0.2\% shorter periods.  In a sense, we use these terms (`A asteroid' and `fragments') as a shorthand to define the orbital periods, rather than as a full acceptance of the physical scenarios they imply.  For convenience, we summarize in Table \ref{tbl:periods} the various periods that we refer to in this work.

\begin{table}
\centering
\caption{WD 1145+017 Reference Periods}
\begin{tabular}{lcc}
\hline
\hline
Description & Period (hr) & Reference \\
\hline
K2 `A' period & 4.4989 & V15$^\dag$ \\ 
`A' period from R16$^*$ & 4.5004 & R16 \\
mean period for fragments & 4.4912 & this work \\ 
derived from BLS transform & 4.4911 & this work \\
ref.~period for 3 waterfall plots$^\S$ & 4.4916 & this work \\ 
ref.~period for Hough transform$^\ddag$ & 4.4950 & this work \\
fragments; waterfall plots$^{**}$ & 4.4900-4.4950 & this work \\
fragments; Hough transform$^{\dag \dag}$ & 4.4903-4.4919 & this work \\
fragments & 4.4905-4.4951 & R16 \\
fragments & 4.4911-4.4951 & G16$^{\ddag \ddag}$ \\
K2 `B' period & 4.6053 & V15 \\
K2 `B' period & 4.6064 & this work \\ 
\hline
\hline
\end{tabular}
\label{tbl:periods}

$^\dag$Vanderburg et al.~(2015); $^*$Also reference period for waterfall plots Figs.~\ref{fig:waterfall} and \ref{fig:waterfall2} in Sect.~\ref{sec:WF}; $^\S$Sects.~\ref{sec:WF} \& \ref{sec:WF2}, Figs.~\ref{fig:waterfall0}, \ref{fig:waterfall00}, and \ref{fig:waterfall3a}; $^\ddag$Sect.~\ref{sec:period}; $^{**}$Sects.~\ref{sec:WF} \& \ref{sec:WF2} and Fig.~\ref{fig:Pfrag}; $^{\dag \dag}$Includes 85\% of the periods; $^{\ddag \ddag}$G\"ansicke et al.~(2016).

\end{table} 

This paper presents 6 months of ground-based optical observations of WD1145, which are a continuation of the 2.5 months of similar observations upon which R16 was based. We describe the new set of observations in Sect.~\ref{sec:obser}.  In Sect.~\ref{sec:ahs} we describe the way in which we represent the lightcurves with an analytic fitting function. In Sect.~\ref{sec:WF} we present ``waterfall'' diagrams (stacked plots of normalized flux vs.~phase) in two formats, and review how ``drift lines'' for dip patterns are used to infer the period of the dust clouds that produce the dips. Section \ref{sec:WF2} focuses attention on the waterfall plots for 2016 April, when two unexpected dust cloud events occurred. In Sect.~\ref{sec:activity} we define a quantitative measure of ``dip activity'', and show how the overall dip activity level in WD1145 has changed during the past two years.  A general search for periodicities in dips is described in Sect.~\ref{sec:period}, including the use of box least squares search, interval matching, and Hough transform algorithms.  Discussion Section \ref{sec:discn} reviews four photometric surprises during one year, presents a case for the requirement that dust production is continual, suggests that a connection must exist between the dust clouds and a debris disk, estimates how many WD stars are like WD1145, and describes an important role for future amateur monitoring of this star. We summarize our results and conclusions in Sect.~\ref{sec:concl}.    

\section{Observations}
\label{sec:obser}

\begin{figure*}
\begin{center}
\includegraphics[width=0.95 \textwidth]{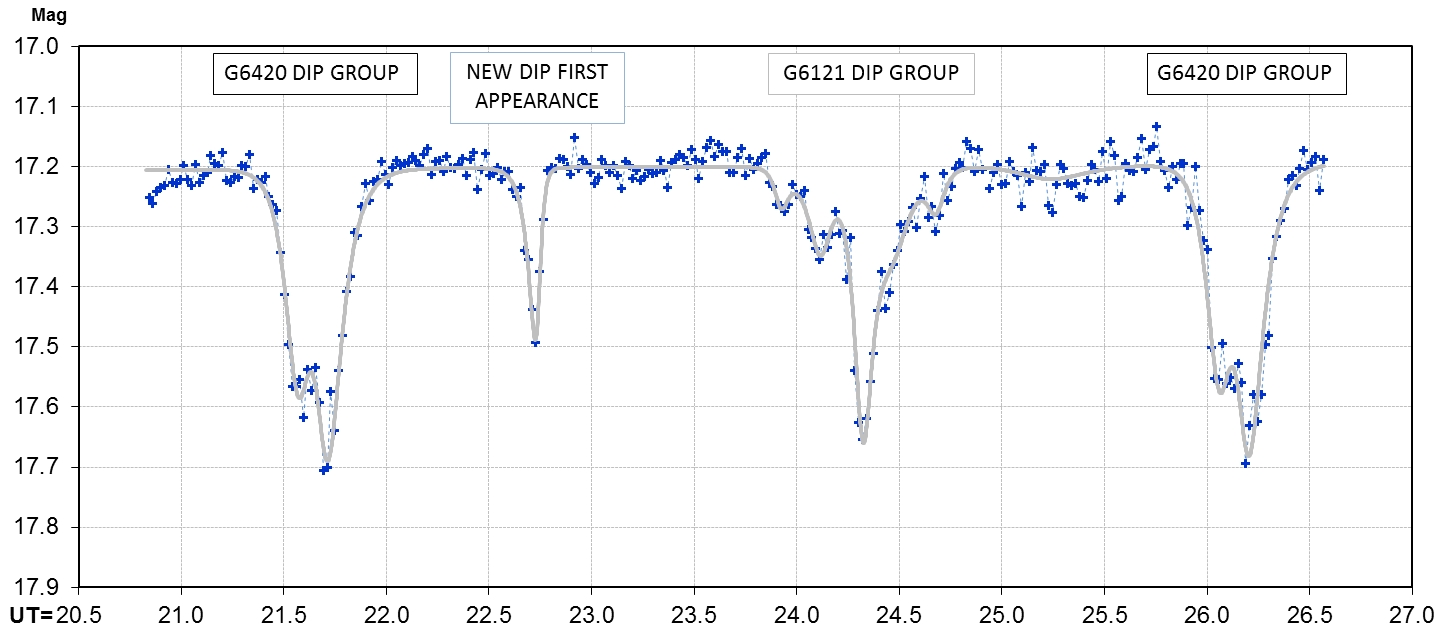}
\caption{Light curve for 2016 April 26, obtained with the IAC80 observatory, covers a 1.3-orbit interval. A group of dips that first appeared on April 20 can be seen twice, at 21.7 and 26.2 UT. The large group of dips that first appeared on about 2016 January 21 is centered at 24.3 UT. A new dip that made its first appearance on this lightcurve is at 22.7 UT. This dip was eventually determined to reappear at 4.6065-hour intervals, which corresponds to the K2 B-period. Names of dip groups are described in Sec. 4.}
\label{fig:lightcurve} % Figure 1
\end{center}
\end{figure*}

We report on 105 observing sessions during the 6-month interval starting 2016 January 25. In addition, some of our results make use of 53 observing sessions from 2015 November 08 to 2016 January 21, that were the basis for R16. Observations for the second observing interval are new, and are the primary source for the present analysis. Since most of the observations for both observing intervals were made by the same observer team, and all were archived and processed in exactly the same way, they will be the subject of this publication without distinction between which observing date interval is being used. The combined data set consists of 158 lightcurves during an 8-month interval. This averages to nearly one lightcurve per day. During each lunar month WD1145 cannot be observed for $\sim$4 days because of the moon's proximity (sometimes occulting WD1145). This high level of coverage is needed for establishing accurate dates for the origin, evolution, and duration of specific events. A brief description of the observatories and data reduction processing follows.

The Gary observations, using the Hereford Arizona Observatory (HAO), used a 14$''$ telescope and were reduced by author BLG. The observatory, observing procedure and reduction process are described in R16.  

The ``Kaye/Gary JBO'' observations were conducted with the Junk Bond Observatory (JBO) 32$''$ telescope by author TGK of the Raemor Vista Observatory. Image sets were calibrated and measured by BLG (hence the ``slash'' in the name referring to these observations). The observatory, observing procedure and image reduction process are described in R16. 

The ``Alonso'' observations were performed using the 0.82-meter (32'') IAC80 telescope on the Canary Islands. It uses the CAMELOT camera with a 2K $\times$ 2K E2V CCD42-40 detector, with a plate scale of 0.304$''$/px. The exposure time was fixed to 60 s, and a 2-channel readout at 500 kHz was used to obtain a final cadence of roughly 68 s. No filter was used. Aperture photometry on WD1145 and 8 reference stars were obtained with custom IDL routines that estimate the centroid of each star through Gaussian fits, compute the flux inside an aperture centered on the star and subtract a sky background annulus value. One of the reference stars was used as a check star to obtain the best combination of reference stars (defined as the one returning the lowest dispersion on the check star). The 4 best stars were typically selected to produce a master reference star with which to compute differential lightcurves. %A total of 5300 frames were obtained on 23 different nights.  

The ``Hambsch'' observations were conducted with a 20" telescope located in Chile, owned by F. J.-H., and remotely controlled from Belgium. It uses a FLI PL16803 CCD camera with a 4K $\times$ 4K Kodak KAF-16803 image sensor. The exposures were 60 seconds and a clear filter was used. Aperture photometry using the freely available software {\tt LesvePhotometry} was used together with reference and check stars. 

\begin{table}
\centering
\caption{Observing Session Count}
\tiny{
\begin{tabular}{lcccccc}
\hline
\hline
Observer &
Gary& 
Kaye/Gary&
Alonso$^\dag$&
Hambsch$^\dag$&
Totals*&
 \\
\hline
Nov&7&1&0&0&13\\ 
Dec&8&9&0&1&21\\ 
Jan&4&0&8&0&19\\ 
\hline
Jan &2&0&1&0&3\\
Feb&12&0&6&0&18\\
Mar&10&0&2&1&13\\
Apr&13&5&7&8&33\\
May&6&12&1&6&25\\
Jun&6&0&0&0&6\\
Jul&7&0&0&0&7\\
\hline
Totals&75&27&25&16&158\\
\hline
\hline
\end{tabular}
}
\label{tbl:decomp}

\scriptsize{*Nov and Dec are for 2015; Jan through June are for 2016. Totals for Nov, Dec and Jan include 16 by Benni and Foote (described in R16). Jan entries are split for < Jan 25 (R16) and > Jan 25 (this work). \\
$^\dag$Some of the Alonso and Hambsch lightcurves were not used for waterfall and dip statistics because of redundancy; but all data are included in the data archives.} 

\end{table}

\section{Fitting Light Curves with Dip Structures}
\label{sec:ahs}

An illustrative lightcurve from our observations for a given night is shown in Fig.~\ref{fig:lightcurve}. That lightcurve was sufficiently long so that more than one 4.5-hour orbital cycle was observed.  The same dip feature seen near the start of the observation was observed again the next orbital cycle toward the end of the observation. The deepest dips on this particular evening were 37\% with typical uncertainties in the flux of $\sim$3\%.  Some specific named dip features, to be discussed in this work, are labeled on the plot.  Later, we stack up a number of these plots, phased to a particular ephemeris, to produce what we call ``waterfall diagrams''.  It will be convenient to use not only the calibrated lightcurve plots, as illustrated in Fig.~\ref{fig:lightcurve}, but also to represent the data as normalized flux and additionally in a symbolic way using fitting functions to describe the dips.

The fitting functions we use are asymmetric hyperbolic secant (`AHS') functions, as described by Eqn.~(1) in R16. Each dip has four parameters: the AHS function's time of minimum, $t_0$, depth, $D$, and the characteristic ingress and egress times, $\tau_{\rm in}$ and $\tau_{\rm eg}$, respectively. Values for these parameters are initially set manually, for all dip features, and automatic refinement is then performed by minimizing $\chi^2$ using the Levenberg-Marquardt algorithm (Levenberg 1944; Marquardt 1963). The set of four AHS parameters for each dip is saved in an archive for later use in creating waterfall plots, investigations of dip properties versus date, and activity level versus date. The dip archive includes 870 entries, each with information about observer, $t_0$, $D$, $\tau_{\rm in}$ and $\tau_{\rm eg}$.

\section{Waterfall Plots for 8 Months of Observations}
\label{sec:WF}

\begin{figure*}
\begin{center}
\includegraphics[width=0.75 \textwidth]{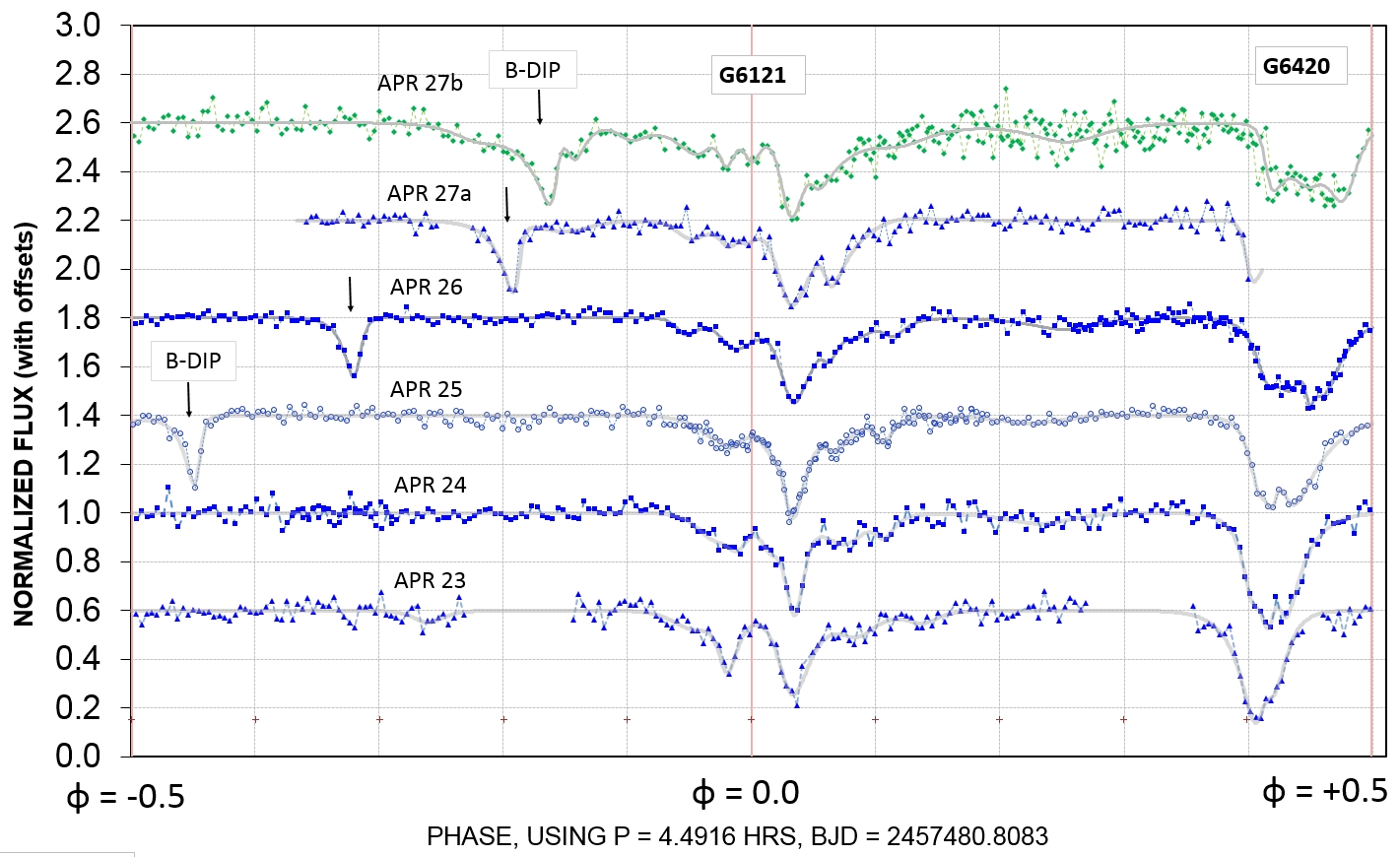} 
\caption{Traditional waterfall diagram spanning 5 days (2016 April 23/24 to April 27/28) using an ephemeris defined by the dust clouds (i.e., fragments). The first 5 lightcurves are from the IAC80 32$''$ telescope and the last one is from a 20$''$ telescope.}
\label{fig:waterfall0} % Figure 2
\end{center}
\end{figure*}

Figure \ref{fig:waterfall0} is a ``traditional'' waterfall plot showing a stack of 6 phased lightcurves spanning 5 days, 2016 April 23/24 to April 27/28, progressing in time in the upward direction. Phase is calculated using a period of 4.4916 hours, and zero phase is defined by BJD = 2457480.8083. This ephemeris (period and phase) was chosen so that zero phase would display a group of dips that appeared on about 2016 January 21, with most of them persisting until the cut-off date of observations for this publication (2016 July 13). During the narrow range of dates in Fig.~\ref{fig:waterfall0} there are two notable events: (1) the sudden appearance of a dip on April 25/26 (3rd lightcurve from bottom, at phase = $-0.45$), which was found to drift to later phases at a rate that corresponds to the K2 `B' period, and (2) a group of dips (at phase about +0.40) that appeared on April 21 (referred to as `G6420') and which evolved rapidly to reveal 3 or 4 distinct dips that drifted slowly to the right in this diagram. 

\begin{figure*}
\begin{center}
\includegraphics[width=0.75 \textwidth]{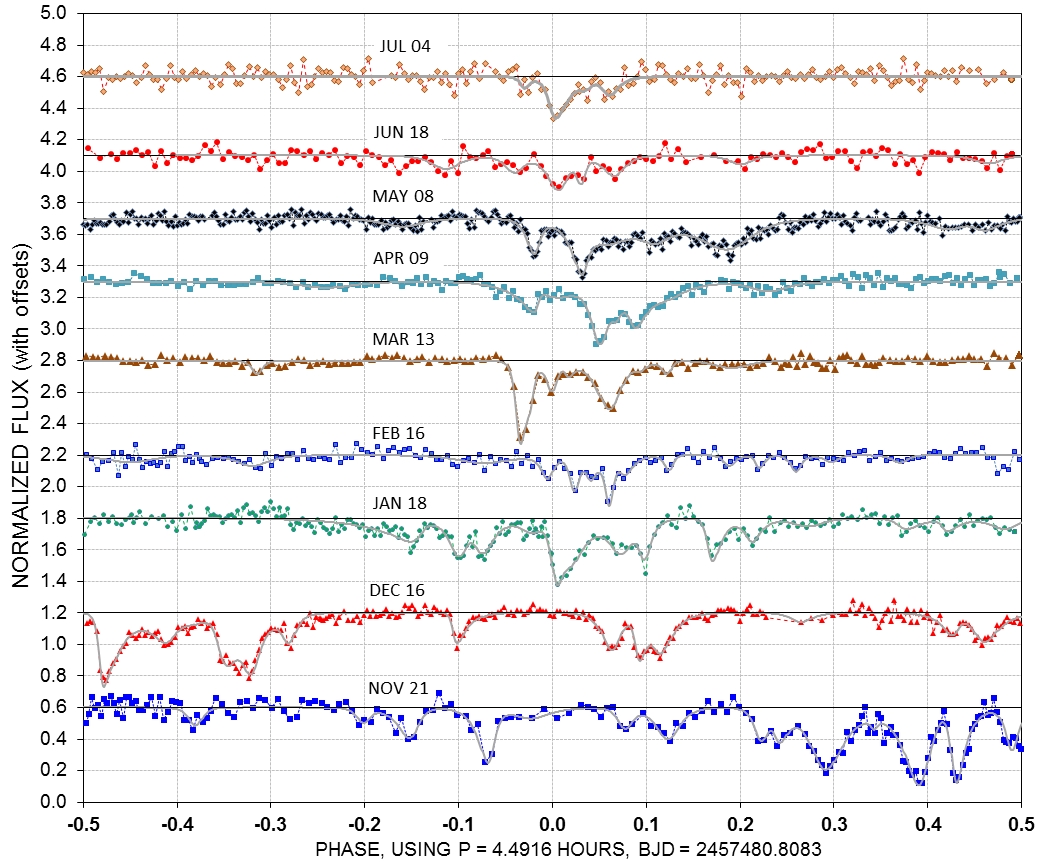}  
\caption{A sampling of phased lightcurves from each month using the same dip drift period ephemeris as in Fig.~\ref{fig:waterfall0}. All observations shown were made by amateurs using backyard observatories (six were made with a 14$''$ telescope, one is a combination of data from 14$''$ and 16$''$ telescopes, and two are combined data from 14$''$ and 32$''$ telescopes). }
\label{fig:waterfall00} % Figure 3
\end{center}
\end{figure*}

In Figure \ref{fig:waterfall00} we show in waterfall format, eight representative lightcurves taken throughout the 8 months of our observations. Note that at all times over the course of the 8 months the dips were sufficiently deep to be accurately recorded by our relatively small aperture telescopes. The first lightcure, in November, has the most dip activity, and the last one shown, in July, has the least. Starting in January most dips are found at the same phase. Whereas the day-to-day lightcurves (see Fig. 2) exhibit similar structures, allowing identification of the same dip in each, which enables drift direction and rate to be measured, on monthly timescales the dip structure is poorly correlated. 

An advantage of this ``traditional'' waterfall format is that it shows dip structure and depth clearly. But a shortcoming is that it cannot accommodate large numbers of lightcurves without becoming very cluttered looking. Also, precise phase locations are not easily determined without a fitting function, and this information is needed for the measurement of drift rates (i.e., orbital periods). 

\begin{figure*}
\begin{center}
\includegraphics[width=0.95\textwidth]{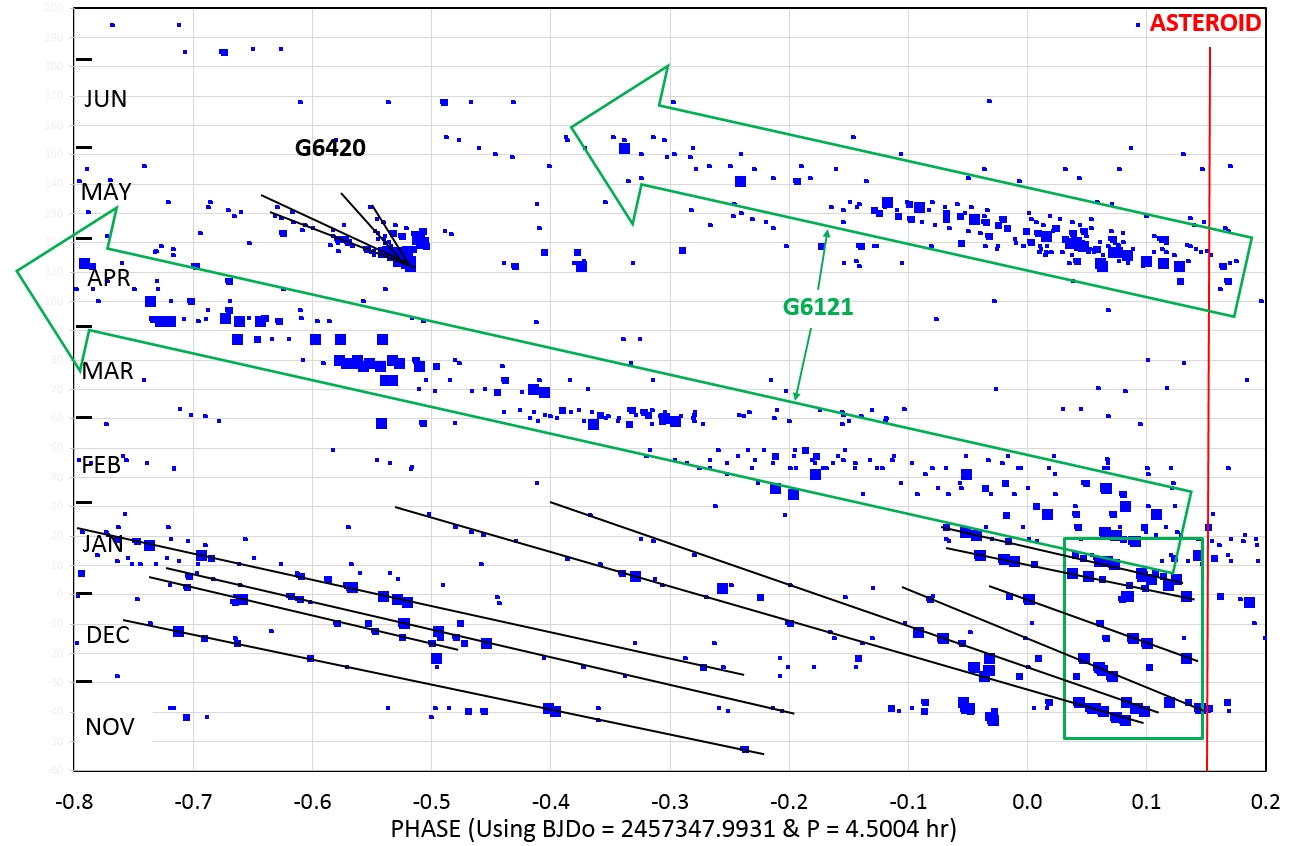} 
\caption{Waterfall plot of all the dips detected during 8-months of monitoring WD1145, as located in date/phase space using an ephemeris defined by the `A asteroid' ephemeris (vertical red line; based on the 2015 Nov/Dec observations within the vertical box). In this type of waterfall plot each symbol marks a dip feature with symbol length corresponding to dip ``scale length'' (duration) and symbol vertical thickness proportional to the dip depth. }
\label{fig:waterfall}  % Figure 4
\end{center}
\end{figure*}

Figure \ref{fig:waterfall} is a waterfall diagram of a different format. Here, each dip is replaced by a bar whose length is the dip scale length, $\tau_{\rm in}$ + $\tau_{\rm eg}$, and whose thickness is proportional to the dip depth (both as determined from the AHS fits). This type of waterfall diagram can accommodate large numbers of lightcurves, covering a long interval of dates; 8 months for this case. It also assigns time to the Y-axis with date increasing in the upward direction. The X-axis is orbital phase, using an ephemeris with BJDo = 2457347.9931 and P = 4.5004 hrs. This ephemeris was not only suggested by the most prominent periodic component of variation in the K2 data, the `A' period, but it was the most prominent component in the first half of the data analyzed in R16. Table 4 in R16 lists an ``A period ephemeris'' epoch corresponding to phase = $+0.04 \pm 0.06$ using the A-asteroid ephemeris given above. As suggested in R16, the group of dips observed during 2015 November and December are from fragments that broke away from the asteroid, and the typical phase marking their presence will be determined by how long they are active in producing dust clouds. In this work we have chosen an ephemeris that intentionally places the A-asteroid crossing in front of the white dwarf star at the apparent origin of this group of dips, which can be seen in Fig. 4 to be at phase = +0.15. 

If the inclination of the asteroid's orbit were close enough to $90^\circ$ so that it crossed in front of the WD, it would produce a transit at times corresponding to the vertical red line.  Since the asteroid is very much smaller than the star, these transits are not expected to be directly observable (e.g., because dips would be much less than 1\%). However, if dust clouds were present close to the asteroid, and if the opaque portions of them were sufficiently large, every time the asteroid passed in front of the WD a transit fade event, or dip, would be observed. Dips would then be confined close to the vertical red line.

Instead of observing this type of pattern, we observe dips that form a pattern of lines branching off the red line and drifting to the upper-left. This was the pattern observed during the entire observational interval covered by the R16 study. R16 explained this using a model in which fragments of the asteroid broke free from near the L1 point (since we hypothesized that the asteroid's Roche lobe had shrunk to the size of the asteroid during an inward migration). Because the fragments went into orbits with slightly smaller semi-major axes they completed their orbits slightly ahead of the asteroid, so any dust clouds that were generated by the fragments would produce transits occurring earlier with each orbit than the asteroid's passage in front of the star. 

Several such ``drift lines'' are shown in Fig.~\ref{fig:waterfall}. R16 identified 15 drift lines, and provided a table of their associated periods. Some drift lines begin close to the red line, and these are presumably caused by fragments that become active in producing dust clouds immediately upon breaking away from the asteroid. Other drift lines begin at phases corresponding to days or weeks later than their hypothetical fragment breakaway date, defined as the date when the drift line projects backward in time to the red line. Drift lines require $\sim$90 days to complete a full phase cycle (i.e., lapping the asteroid). With this model, the explanation for a drift line appearing $\sim$45 days after a presumed breakaway date, for example, is that it remained ``dormant'' for $\sim$45 days before it began to produce a dust cloud (i.e., sufficiently large and opaque to produce observable dips). 

The vertically oriented box in Fig.~\ref{fig:waterfall}, located on the left of the vertical red line, includes many of the dips prior to mid-January. According to the model just described, these dips would be fragments that became active soon after breaking away from the asteroid. Since most of  them had reduced activity by the time they drifted about 0.1 phase units to the left there is a fall-off of dip activity for phases $\lesssim - 0.06$. This pattern accounts for the R16 detection of a strong periodicity at the K2 `A' period.

Something significant in the waterfall diagram changed on about January 21 (it is just a coincidence that this is the cut-off date for observations that were included in R16). On the assumption that the breakaway model described in R16 is correct, on this date a group of fragments producing a large number of dips appears to have broken away.  The fragments in this group produced dips with a group drift pattern which lasted for the 6 months that remained for the observational coverage of this work. This group of dips will be referred to as G6121 (i.e., G = group, 6 = 2016, 1 = January, 21 = 21st day; note: months are given in hexadecimal). Their drift rate corresponds to a period of $\sim$4.4912 hrs (see also Croll et al.~2015). Another interesting thing to note is that when G6121 appeared there was a cessation of new individual breakaway events.  

\begin{figure*}
\begin{center}
\includegraphics[width=0.70 \textwidth]{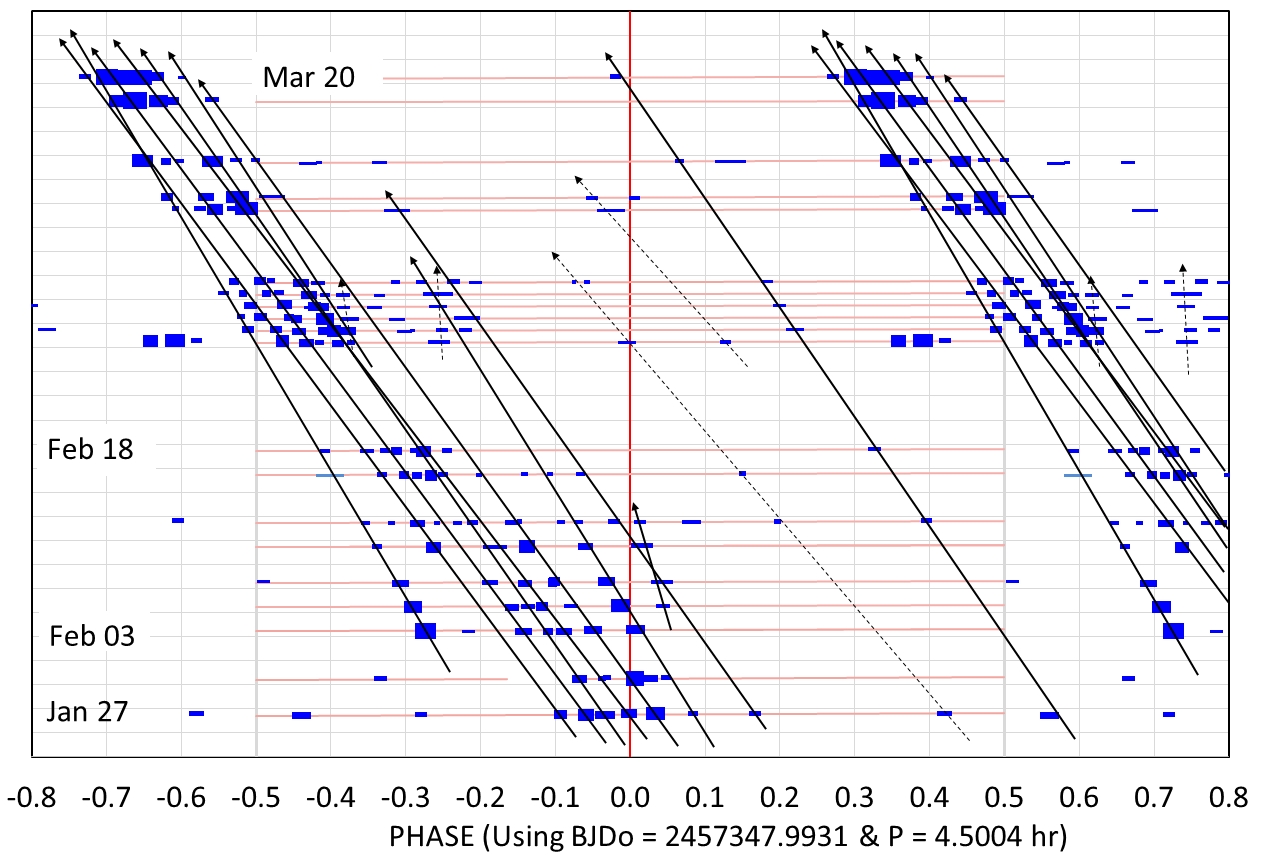} 
\caption{Detail of waterfall diagram shown in Fig.~\ref{fig:waterfall} for the $\sim$2-month interval centered on 2016 February (January 27 to March 20).  The  drift lines connecting the dips are suggestive and ambiguities may arise when dip features cross.}
\label{fig:waterfall2} % Figure 5
\end{center}
\end{figure*}

It is not clear from Fig.~\ref{fig:waterfall} that any distinct drift lines exist for the dips within G6121 since the dip symbols are too close together to see such a pattern. Fig.~\ref{fig:waterfall2} is an expanded version for a 2-month date range starting with the G6121 creation date. It shows that many of the dip features may plausibly be associated with distinct drift lines. In other words, these dip features are from dust-cloud producing fragments that appear to last several months.   

Referring back to Fig.~\ref{fig:waterfall}, in late April, at phase $-$0.5, a group of dips appear, labeled `G6420'. The 3 or 4 drift lines for this group  have different slopes, and all of them are rotated clockwise with respect to the G6121 slope, signifying that they have longer periods than the long-lived G6121 group. The G6420 group is treated in the next section.	

\section{April/May Waterfall Diagram}
\label{sec:WF2}

\begin{figure*}
\begin{center}
\includegraphics[width=0.60 \textwidth]{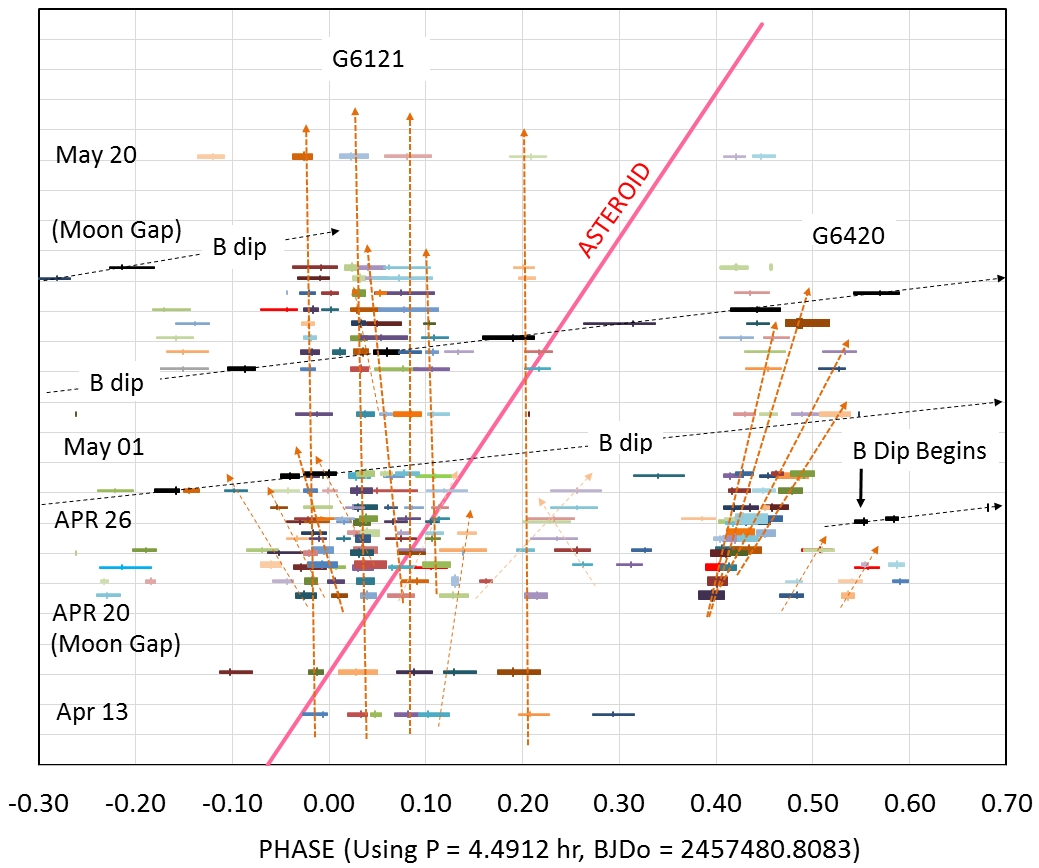} 
\caption{Waterfall diagram for the $\sim$40-day interval centered on early 2016 May (April 13 to May 20). The red line shows the location of the `A' asteroid. The different symbol colors have no significance other than helping to distinguish one dip from neighbor dips which sometimes overlap.}
\label{fig:waterfall3} % Figure 6
\end{center}
\end{figure*}

\begin{figure}
\begin{center}
\includegraphics[width=0.476 \textwidth]{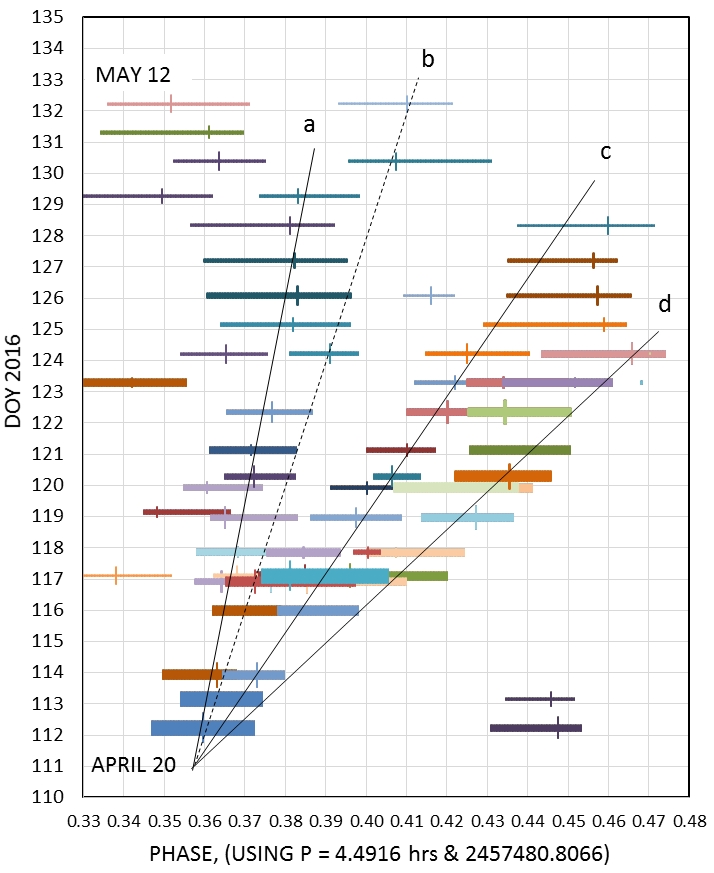} 
\caption{Enlargement of the previous waterfall plot, showing the G6420 group in greater detail. Symbol colors have no significance aside from the explanation given in the previous figure.}
\label{fig:waterfall3a} % Figure 7
\end{center}
\end{figure}

\begin{figure}
\includegraphics[width=0.46 \textwidth]{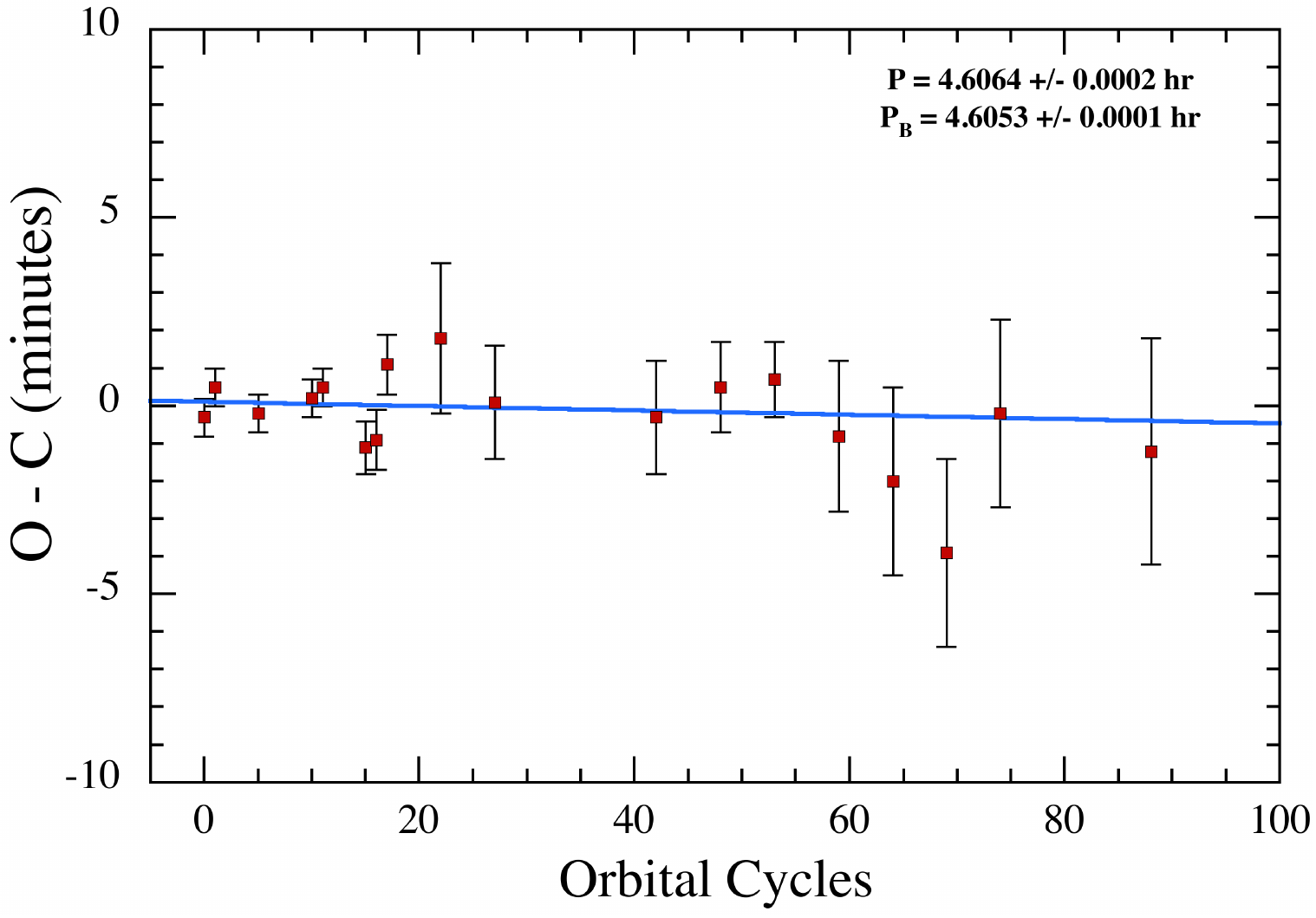}
\caption{Difference between observed and model-predicted times (``O-C'') for the ``B dip''.}
\label{fig:Bperiod} % Figure 8
\end{figure}

Figure \ref{fig:waterfall3} is an expanded version of the Fig.~\ref{fig:waterfall} waterfall diagram for the 50-day interval, 2016 April 10 to May 30, using a slightly different display format. It uses an ephemeris suggested by an initial analysis of the G6121 drift rate (P = 4.4916 hrs). By using this period these drift lines appear close to vertical, which makes it easier to identify dip associations. The drift lines near zero phase are long-lasting, and many of the drift lines are nearly parallel. The red line is the `A' asteroid's position (based on the  ephemeris described in the previous section). 

The remarkable G6420 group is more clearly seen in Fig.~\ref{fig:waterfall3a}, showing just the G6420 region of the waterfall diagram. Four drift lines are identified, and they all abruptly begin on April 20, phase = +0.36. This group of dips illustrates what we suggested might be possible in R16, namely that the date for the simultaneous creation of a group of fragments (possibly due to a collision) can be determined by backward extrapolation of drift lines to a date where they converge. In this case, however, no backward extrapolation is required, nor is a creation date conjecture necessary. Even if observations had not been made until a few days after this group's first appearance a backward projection would still have yielded a fairly accurate creation date. 

The drift lines can be referred to with letters appending their group name, e.g., G6420a, b, c and d. Dip `d' fades to unobservable status first, after 13 days. The other dips last longer, especially drift line ``a'' whose dips are clearly present for at least 17 days. Dip `d' has a slope requiring a period almost as long as the `A'-asteroid's period. If the `d' fragment's orbit is eccentric it could conceivably impact the `A' asteroid. The existence of ``b'' is the least certain because its dip features are close to the ``a'' dips and there are only 4 occasions that they were identified as a separate dip feature.

Figure \ref{fig:waterfall3} shows the abrupt appearance of a dip on Apr 26 and phase +0.55. It is very brief, but deep. This is the same dip shown in the Fig. 1 lightcurve at 22.7 UT, and also in Fig. 2 (starting with the 3rd lightcurve from bottom). The dip was found to move to greater phases at a high rate, corresponding to a period of $4.6064 \pm 0.0002$ hrs, as shown in Fig.~\ref{fig:Bperiod}. We refer to this dip as the ``B dip'' because it has a period almost exactly the same as the Kepler K2 `B' period of 4.6053 hours (Vanderburg et al. 2015). Our interpretation of this dip feature is that it is either a fragment that broke off an asteroid having a period close to the K2 `B' period, or it is the `B' period asteroid itself that is producing a dust cloud. 

\section{Activity Based on Light Curve Data} 
\label{sec:activity}

One way of characterizing the `activity level' of a dip is a quantity which we define as the ``equivalent width'' (`EW'), i.e., the integral under the dip curve, which we approximate as the product of the depth, $D$, times the sum of the ingress time, $\tau_{\rm in}$, and egress time, $\tau_{\rm eg}$, multiplied by a numerical factor of $\pi/2$, where the latter factor is needed to produce the correct result when using the AHS model (see Sect.~\ref{sec:ahs}):
\begin{equation}
{\rm EW} ~\equiv ~ \frac{\pi}{2}\, D ~(\tau_{\rm in} + \tau_{\rm eg}) \,P_{\rm orb}^{-1}
\label{eqn:EW}
\end{equation}
Here we have normalized EW by an orbital period of $\sim$4.5 hours so as to make EW dimensionless; also, EW has values between zero and one. If the portion of the dust cloud that transited in front of the WD were optically thin, then EW would be proportional to the projected area of all dust particles (assuming that their circumferences are greater than $\sim$ one wavelength, i.e., approaching a scattering cross section roughly equal to the geometric area of the grains). In particular, one can show that EW $\simeq A_g/(4 \pi R_{\rm wd} d)$, where $A_g$ is the total grain area passing in front of the WD, $R_{\rm wd}$ is the radius of the WD, and $d$ is the orbital distance of the dust cloud.

\begin{figure}
\begin{center}
\includegraphics[width=0.49 \textwidth]{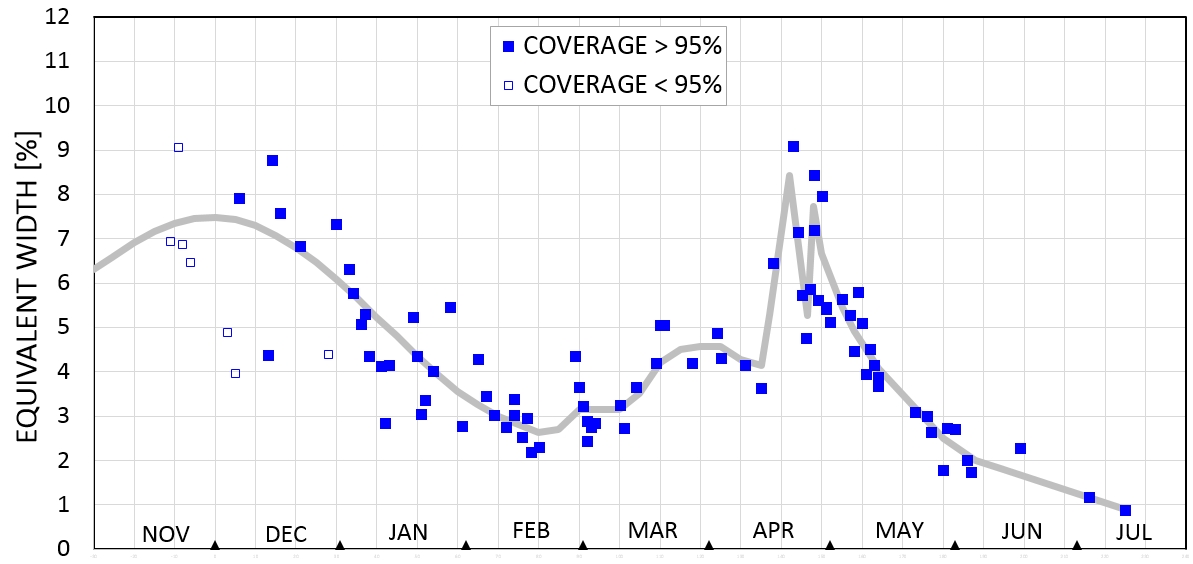} 
\caption{Overall dip activity vs. date during the 8-months of observations (2015 November 08 to 2016 July 13). Equivalent width is defined in the text.  }
\label{fig:activity1}  % Figure 9
\end{center}
\end{figure}

We can use the fitted AHS parameters for all of the lightcurves to derive a measure of the total source `activity' vs.~time over the course of our 8 months of observations. We accomplish this by computing the EW of the lightcurves (i.e., the ``area under the curve'', as defined above). Hereafter, we refer to this definition of ``equivalent width'' as ``activity level''.  Figure \ref{fig:activity1} shows ``equivalent width'' dip activity versus date for the 8-month interval of this study. Activity begins at a high level in 2015 November, decreases to a low value in mid-February, rises to a similar high level in late April, after which it undergoes an apparent monotonic decrease that continues until the last observation date in July.

\begin{figure*}
\begin{center}
\includegraphics[width=0.7 \textwidth]{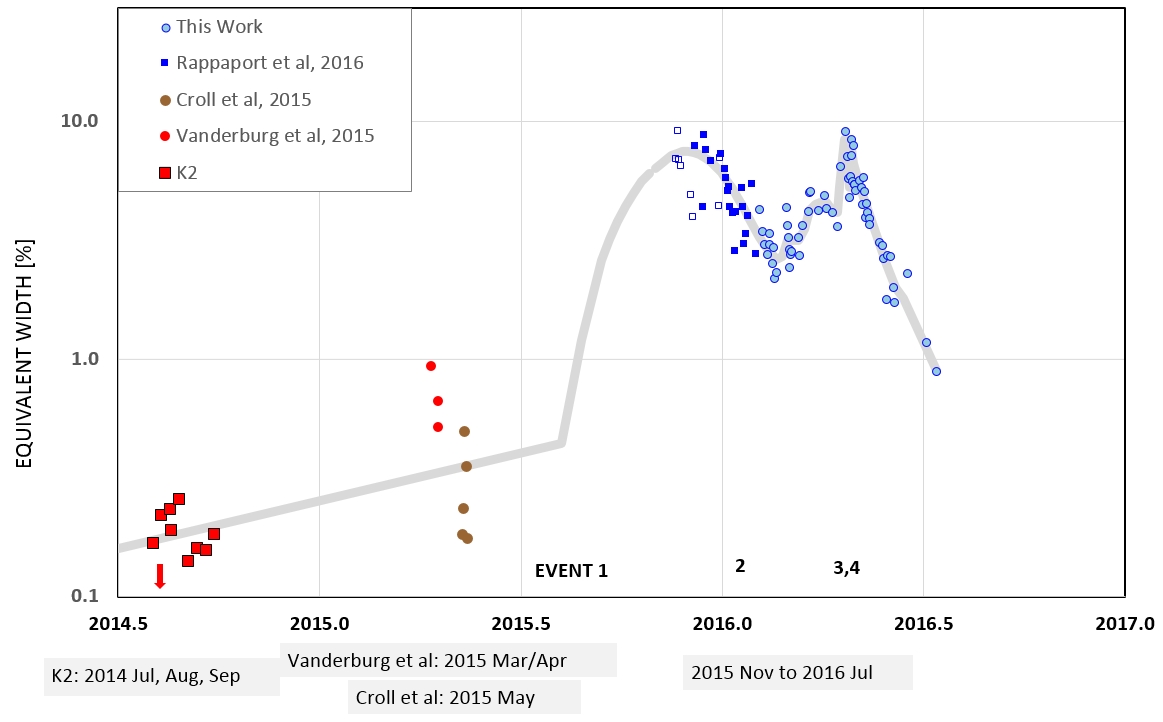} 
\caption{Overall dip activity, expressed as an EW (see Eqn.~\ref{eqn:EW}) vs. date, from 2014 ({\em Kepler} K2) to 2016 July 13.}
\label{fig:activity2}  % Figure 10
\end{center}
\end{figure*}

Figure \ref{fig:activity2} shows the same data on a log scale, for a 2-year interval in order to include the {\em Kepler} K2 observations and follow-up ground-based observations reported by Vanderburg et al.~(2015) and Croll et al (2015). As described in R16, the {\em Kepler} K2 discovery observations, as well as the follow-up ground-based observations in 2015, were made when the ``dip activity'' was at a low level. When amateurs started observing in 2015 November the dip activity level was $\sim$25 times greater. This dramatic rise in activity must have occurred between June and October of 2015.  

In studying all the dip features for WD1145, we have seen that, thus far, no dips deeper than 60\% have been measured. Either the dust clouds have not been opaque enough to produce deeper dips, or they are opaque and have not covered more than 60\% of the stellar disk of the WD. These two extreme cases can be thought of as ``optically thin'' and ``optically thick'' models. Knowing which of these alternatives to model, or what combination of them to employ, will be part of any analysis attempting to quantify dust production and loss mechanisms. The shape of the dips can sometimes differ for the two alternatives (see discussion in Alonso et al.~2016). Many dips have begun with sharp structures (e.g., $\sim$3 minutes wide and $\sim$50\% deep), but none of them in the present archive can be characterized with certainty to be ``flat-bottomed'', as would be expected in a hard-body transit. Detailed modeling of dip structures will be justified when larger aperture telescopes produce an archive of lightcurves with better SNR than those in the present archive.  

\begin{figure}
\begin{center}
\includegraphics[width=0.47 \textwidth]{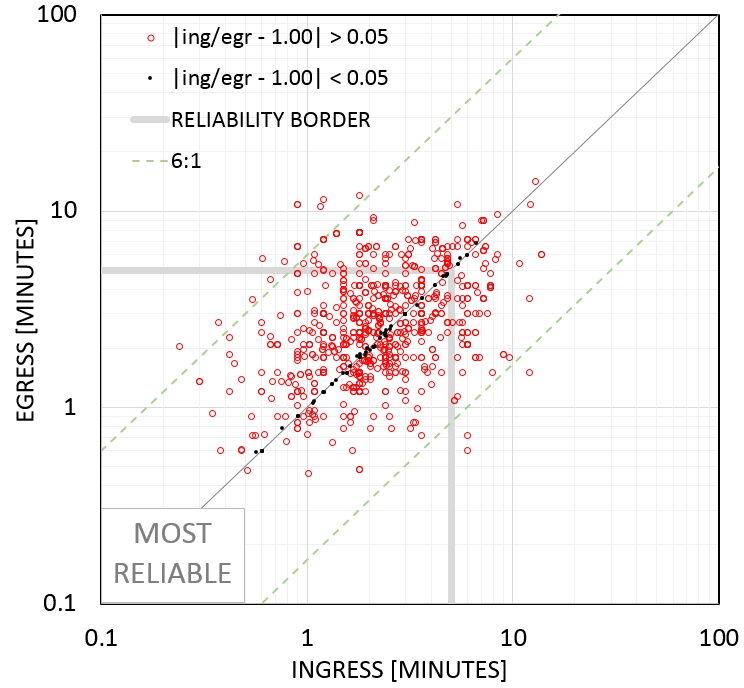} 
\caption{$\tau_{\rm eg}$ vs. $\tau_{\rm in}$ for 622 dip solutions over the course of 7 months.  Only dips with depths $D > 0.05$ are included (D has units of normalized flux). The red circle symbols are for tau ratios either < 0.95 or > 1.05; the black dots are for ratios between 0.95 and 1.05. In spite of the apparent near symmetry of the two populations of red symbols with respect to the $\tau_{\rm eg} = \tau_{\rm in}$ line, there are actually 360 points with tau ratio > 1.05 compared with only 262 < 0.95, a clear 4-$\sigma$ asymmetry in favor of longer egress times than ingress times. The box corresponds to both $\tau$ values being < 5 minutes, where AHS solutions are most reliable.}
\label{fig:taus}  % Figure 11
\end{center}
\end{figure}

We note that one interesting statistic that can readily be formed from all the measured dips, regardless of which drift line they belong to, is a correlation between the dip egress times and the ingress times.  We show in Fig.~\ref{fig:taus} such a correlation. The first thing we see from this figure is that the median values of both $\tau_{\rm in}$ and $\tau_{\rm eg}$ are about 2.2 min.  The `duration', $\Delta t$, of the dip can be expressed in terms of the parameters of the AHS model as:
\begin{equation}
\Delta t \simeq 2 \,(\tau_{\rm in}+\tau_{\rm eg})
\end{equation}
from which we conclude that the median dip duration is about 9 min.  Finally, we note that for a $\tau$-ratio defined to be $\tau_{\rm eg}/\tau_{\rm in}$ there are somewhat more data points in Fig.~\ref{fig:taus} with high $\tau$-ratios (>1.05) than low $\tau$-ratios (<0.95), indicating a slight tendency for egress times to be longer than ingress times.  In fact, there are 360 high $\tau$-ratio points vs. only 262 low $\tau$-ratio ones. This is significant at the 4-$\sigma$ level, and we take it to be a real effect (see also R16).

\vspace{0.3cm}

\section{Period Search}
\label{sec:period}

Thus far our searches for periodicities in the data set have largely been done by direct inspection of the waterfall diagrams. Associating a series of dips and calling them a drift line was done ``by eye'' - i.e., subjectively.  R16 also addresses this subjective procedure, and a list of claimed periods is given for dates before 2016 January 25. However, whereas the lightcurves dealt with in R16 consisted of dips that usually did not overlap, and could be identified as they shifted in phase from dates with separations of two or three days, the lightcurve observed after late 2016 January (the G6121 group) consist of dips that are ``bunched up'' in phase and exhibit frequent and persistent overlaps.  This has made it difficult to identify associated dips for creating drift lines in the post-R16 data. Consequently, our ``eyeball'' drift lines after late January suffer from greater uncertainty, and we are therefore handicapped in assessing the longevity of individual dust cloud activity.  It is our subjective opinion that the data after 2016 January include at least a half dozen drift lines that last 2 - 3 months, and several more that last at least  one month. Longer lasting drift lines are simply too difficult to confirm due to confusion caused by the overlapping of dips with approximately the same phase but slightly different orbital periods. Nevertheless, the G6121 group of many dips endure for about 5 months since their appearance in late January, and during June and July the G6121 component dips gradually disappear. For the individual drift lines we can only be certain of much shorter lifetimes.  The task of associating dip features with the same dust cloud under these conditions is a challenge deserving of more effort than has been possible in preparing this paper.  Since we have data exchange files for anyone who requests them, we invite anyone interested in pursuing this task to do so with our data.

\begin{figure}
\begin{center}
\includegraphics[width=0.48 \textwidth]{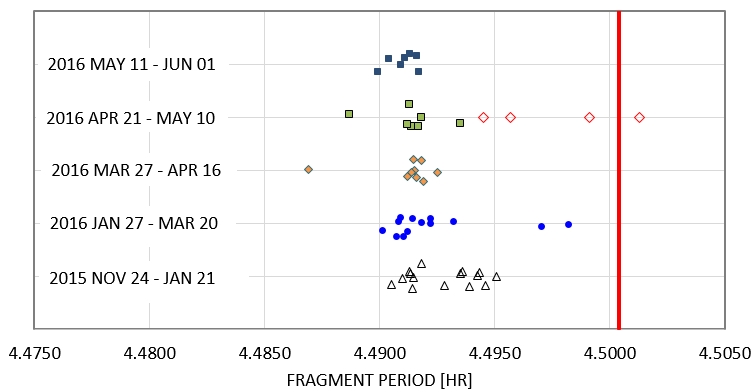}
\caption{Periods of dip features based on drift line slopes during 8 months of observations. }
\label{fig:Pfrag} % Figure 12
\end{center}
\end{figure}

The periods that we have tentatively identified from the drift lines are shown in Figure \ref{fig:Pfrag} and are based on drift line slopes for the entire 8-month set of observations (i.e., shown in Fig.~\ref{fig:waterfall}).  However, note that we have broken up the data into segments of several weeks in order to ensure that we are assigning the correct dip features to a given drift line. We find that all the periods in this plot lie between 4.4870 and 4.5012 hours, a range of only 0.3\%.  This corresponds to orbital radii that are the same to within 0.2\%.  If we remove the 6 periods farthest from the mean, then the remaining $\sim$85\% of the periods all lie between 4.490 and 4.496, a range of only 0.13\%, corresponding to orbits within an extraordinarily narrow ring of only 0.1\% in width.  

\begin{figure}
\begin{center}
\includegraphics[width=0.480 \textwidth]{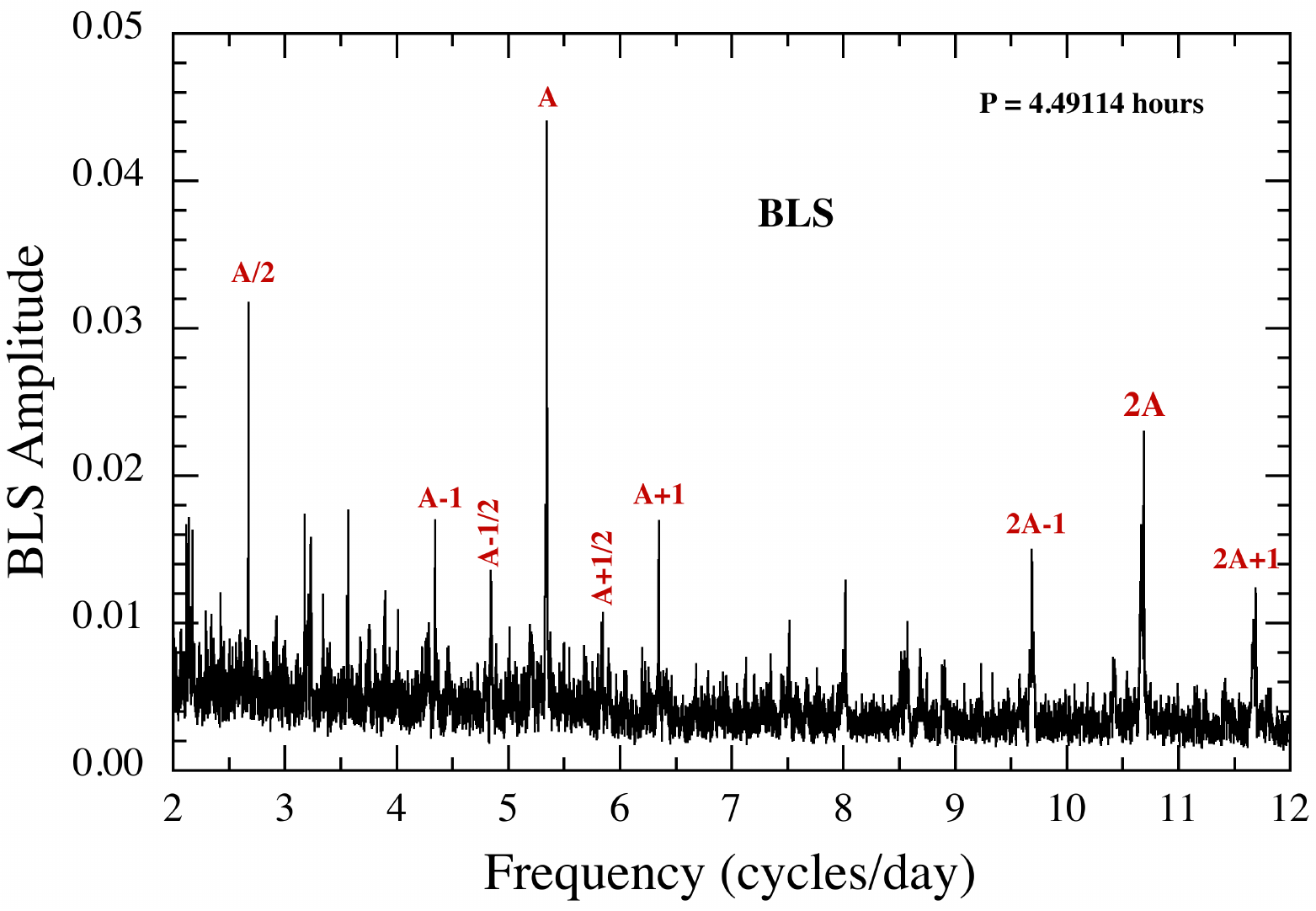} 
\includegraphics[width=0.480 \textwidth]{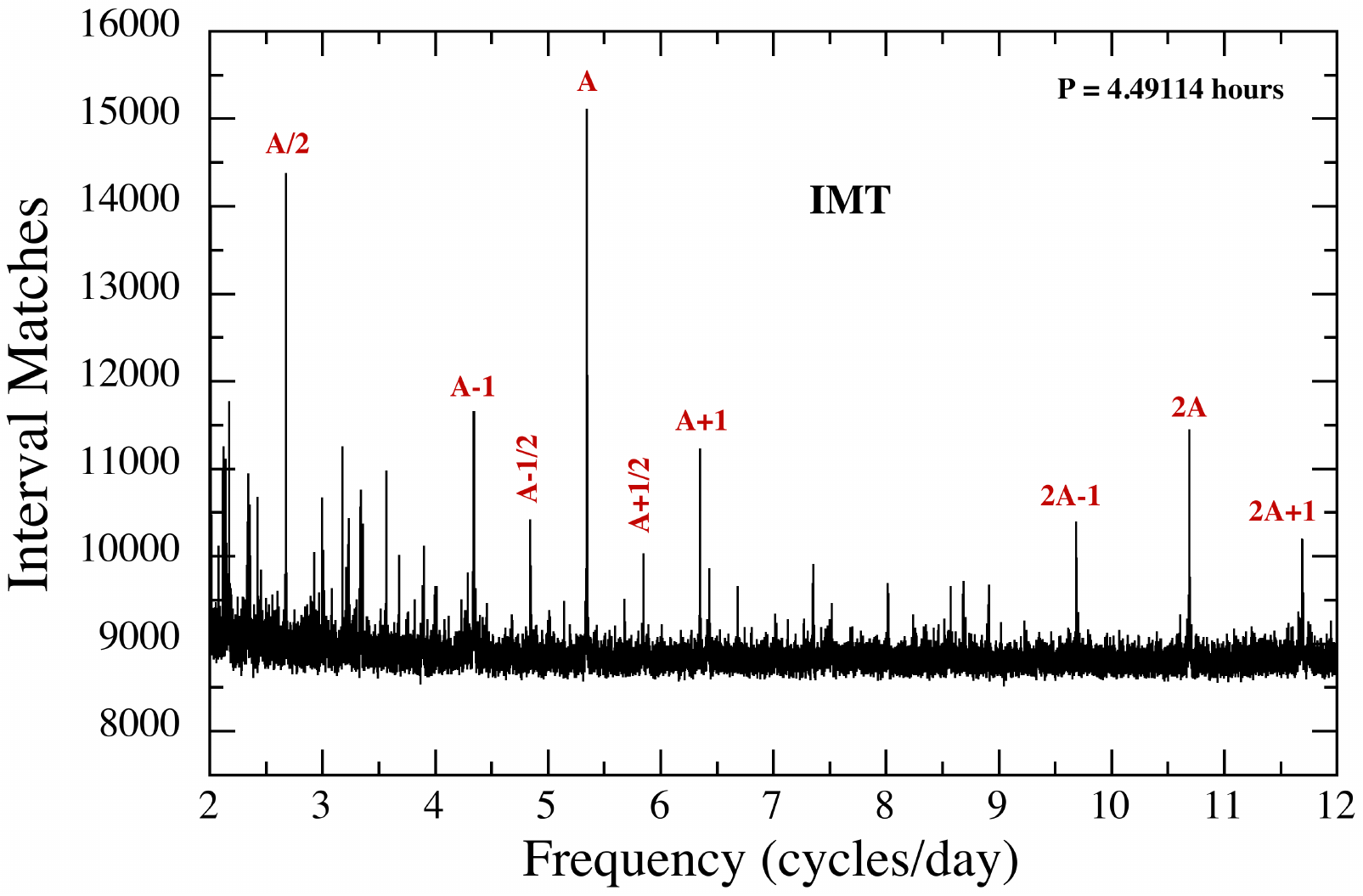} 
\caption{Upper panel: BLS transform of the photometric data from a 4-month observation interval covered by this work (2016 January 27 to 2016 May 20). Bottom panel: Interval Match Transform (see text for definition) for the same observation interval. The dominant period is 4.49114 hours.  All other significant peaks in these transforms are at multiples of the base frequency, or at 1- or 2-day sidebands of the observing window.}
\label{fig:BLS} % Figure 13
\end{center}
\end{figure}

Because of the limitations caused by crowded drift lines we have pursued three other approaches to quantitatively assess the periods that are present, and the longevity of the dip features that produce them.  In the first, we subjected the data to a Box Least Squares (BLS) transform (Kovacs et al, 2002).\footnote{\url{http://exoplanetarchive.ipac.caltech.edu/cgi-bin/Periodogram/nph-simpleupload}}  The results are shown in the top panel of Fig.~\ref{fig:BLS}. What we see is a prominent peak with a frequency of 5.34385 cycles/day, with a frequency resolution of 0.0083 cycles/day, but with an uncertainty in determining the peak frequency of only 0.00042 cycles/day.  This corresponds to a period of $4.49114 \pm 0.00036$ hours. All of the other highly significant peaks in the BLS transform are associated with either this frequency, its higher harmonics, subharmonics, or their beat frequencies with the 1-day observing cycle. The width of this peak, equal to the frequency resolution of 0.0083, is insufficient to resolve the individual periods plotted in Fig.~\ref{fig:Pfrag}.  

The second period search consisted of brute force testing of $10^5$ trial periods against our fitted times of dips (see Sect.~\ref{sec:ahs}). For each trial period we checked whether the time interval between each distinct pair of dips was equal to an integer number of trial periods.  We allowed for a plus or minus 2\% leeway in terms of matching an integral number of cycles.  The bottom panel in Fig.~\ref{fig:BLS} shows the number of ``matches'' for each trial frequency.  This transform, which we call ``Interval Match Transform'' (`IMT'), bears a remarkable similarity with the BLS transform, and even has a higher signal-to-noise ratio at the peak at 4.49114 hours (the same period determined from the BLS analysis). However, here, as with the BLS transform, there is insufficient frequency resolution to separate the individual periods cited in Fig.~\ref{fig:Pfrag}.

Our third approach for determining periodicities in the data involves the use of a Hough transform (`HT'; Hough 1959, 1962;  Duda \& Hart 1972).  The HT was initially specifically designed to look for linear features in two dimensional images, and should be suitable for the detection of drift lines in our waterfall diagrams.  We cast the waterfall data in the form of a discrete array of 200 orbital-phase bins in the ``$\phi$''-direction, and 220 days to span all the observations, in the time ``$T$''-direction.  We use only dips that are deeper than 5\%, i.e., quite statistically significant, and then weight them all equally.  This allows for the possibility that dip features along a given track are likely to vary in strength.  Each point in the domain of the waterfall plot, $W(\phi,T)$, is transformed into a curve in Hough space, $H(\rho,\theta)$ via the expression:
\begin{equation}
\rho = \phi \cos \theta + T \sin \theta 
\label{eqn:hough}
\end{equation}
where $\theta$ and $\rho$ specify the slope and the perpendicular distance from the origin, respectively, of a potential linear feature in $W(\phi,T)$. 

The value of the Hough transform is that all points along a common linear feature in the waterfall diagram lead to a common intersection point $(\theta, \rho)$ in Hough space, $H(\theta, \rho)$. The more points that are accumulated in a given pixel in Hough space due to intersecting curves given by Eqn.~(\ref{eqn:hough}), the more likely that point is to represent a linear feature in the waterfall diagram.  When a `bright' intersection point is found, the angle $\theta$ provides the slope (or tilt angle) of the drift line in the waterfall diagram which, when combined with the reference fold period, yields the corresponding orbital period.

\begin{figure}
\begin{center}
\includegraphics[width=0.465 \textwidth]{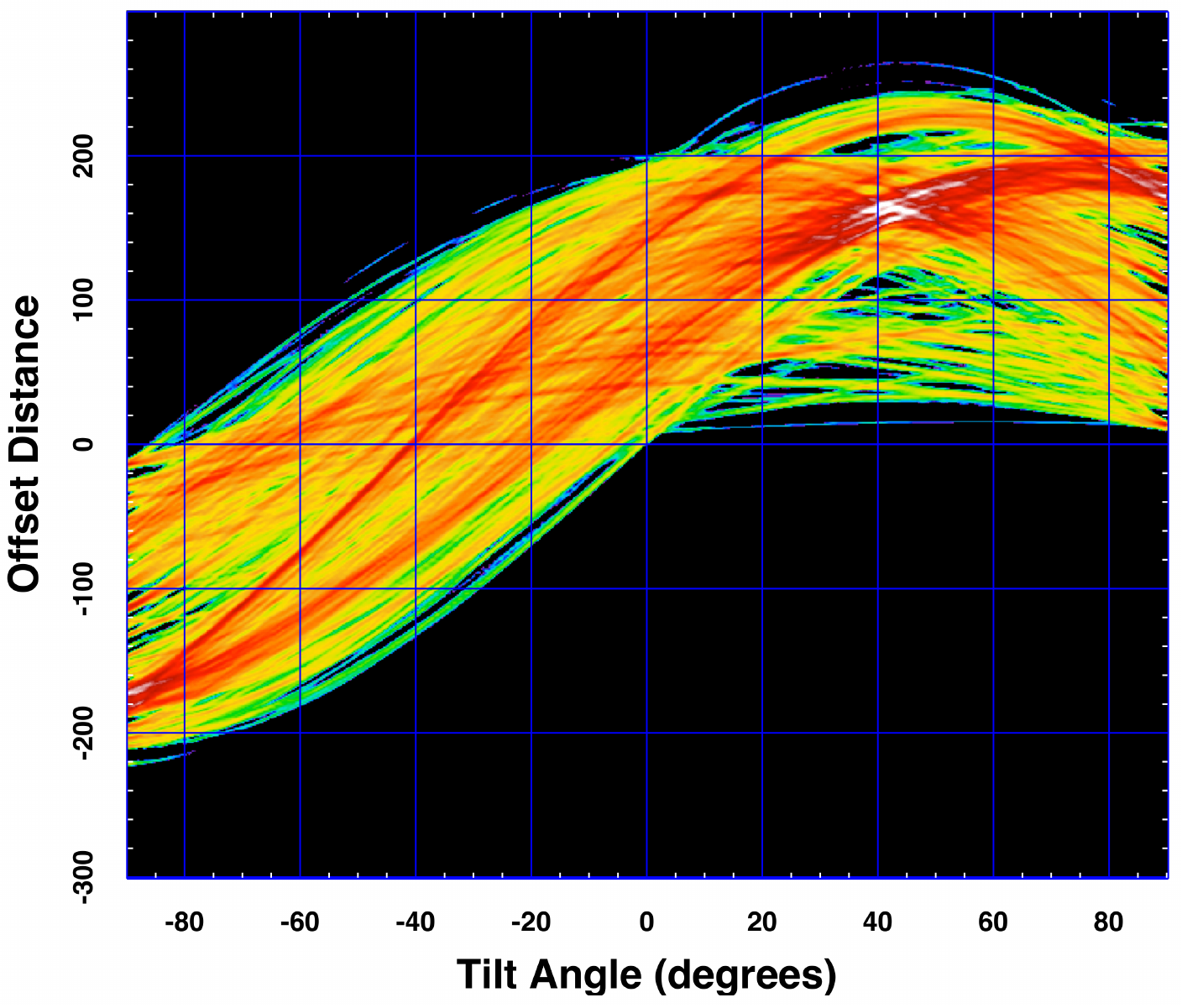} \vglue0.2cm 
\includegraphics[width=0.47 \textwidth]{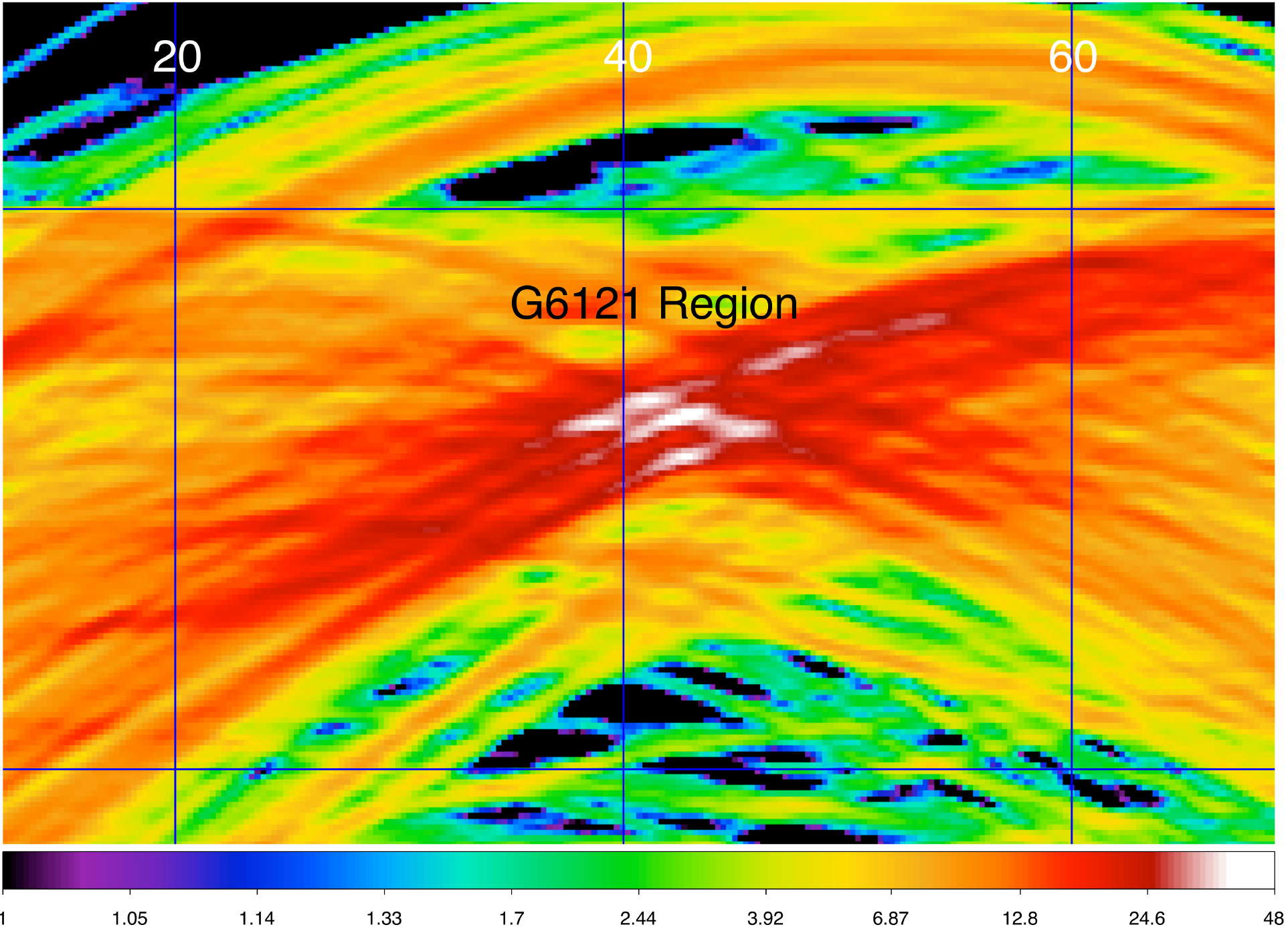}
\caption{Hough transform of the full waterfall diagram from all 8 months of our observations.  The waterfall diagram was produced with a reference fold period of 4.4950 hours, and all dips with depth $>0.05$ were included and given equal weight. The top panel shows the full Hough transform as a function of the drift line tilt angle $\theta$, and offset, $\rho$ (see Eq.~(\ref{eqn:hough})).  The bottom panel is a zoom in on the region corresponding to the substantial number of largely parallel drift features seen in the waterfall plot (Fig.~\ref{fig:waterfall}) with orbital period = 4.4911 hours (the G6121 drift lines). The colors encode the number of dips involved in the linear features that are detected.   The white peaks range over $36^\circ - 48^\circ$, corresponding to a range of periods from 4.4903 to 4.4919 hours. This is the region wherein most of the periods plotted in Fig.~\ref{fig:Pfrag} lie.}
\label{fig:hough} % Figure 14
\end{center}
\end{figure}

The results are shown in Fig.~\ref{fig:hough}.  The waterfall input data were produced with an arbitrary reference fold period of 4.4950 hours \footnote {The reference period was chosen to be near the high end of the fragment periods (see Table \ref{tbl:periods}), so as to avoid having the drift lines run vertically in the waterfall diagram.  This is done for aesthetic purposes only.}. In general, horizontal displacements correspond to different slopes of the drift lines, and hence different orbital periods, while vertical displacements represent the same slopes (i.e., same periods), but different orbital phases for drift lines. The colors encode the approximate number of dips involved in the linear features that are detected.  The region of white spots corresponds to the substantial number of largely parallel drift features with orbital period near 4.4911 hours (i.e., the G6121 features seen in Figs.~\ref{fig:waterfall} and \ref{fig:waterfall2}). The corresponding slopes indicated on the Hough transform (within the white region) range over $42^\circ \pm 6^\circ$, corresponding to a range of periods from 4.4903 to 4.4919 hours. This is just the region wherein most of the periods plotted in Fig.~\ref{fig:Pfrag} lie.  

Overall, the Hough transform has an advantage over the BLS, and IMT transforms by virtue of the fact that it can separate drift lines of the same period but different phases.  However, ambiguities in terms of which dips go with which drift lines still remain.  Another important advantage of the HT diagram is that it is a completely objective analysis tool for identifying drift patterns in a waterfall plot, and it shows that the drift lines derived using more subjective methods provide an accurate evaluation of dip drift rates, and hence periods. In addition, it supports these more subjective analyses that call for longevity values of several months in at least a number of the drift features. 

\section{Discussion}
\label{sec:discn}

\subsection{Four Surprise Events}
\label{surprises}

The 8-months of observations of WD1145 ending on 2016 July 13 (the cut-off date for observations included in this work) have led to four significant surprises: (1) A dramatic increase in ``activity'' occurred sometime between 2015 June and October, causing dip levels to be $\sim$25 times greater than during the K2 observations of one year earlier as well as the follow-up ground-based observations during the first half of 2015. (2) In mid-January of 2016 the dip pattern of activity changed from a steady production of short-lived dips originating at the presumed `A'-asteroid location to a cessation of such activity; at the same time, a large number of longer-lived dips appeared in an inner `A'-fragment orbit. (3) On April 21 a group of 3 or 4 dips appeared at a phase that we interpret as corresponding to a location far from the `A'-asteroid. These dips had a relatively large range of periods, and their drift lines radiated from their origin date. This demonstrates that when drift lines are observed long after the origin of the fragments that eventually produce dust clouds and observable dips, it is still possible to infer a creation date by projecting drift lines backward in time to a convergence date of presumed origin. (4) On April 26, 2016 a dip suddenly appeared that repeated with a period corresponding to the K2 `B'-period, with P = 4.6064 hours. This validates the existence of an asteroid at the `B'-period orbit, and provides indirect support for the existence of the `C' through `F' asteroids. We assert the latter simply by virtue of the fact that heretofore none of the other five periodicities claimed from the K2 observations have been confirmed from ground-based observations. 

\subsection{The Case for Quasi-Continual Dust Production}
\label{sec:continuous}

Whenever observing sessions were long enough to observe more than one 4.5-hour orbital cycle, we always observed a near repeat of the dip pattern from the previous orbit (see, e.g., Fig. 1). That is, changes in the dips typically take numerous orbits, or even days to evolve.  Some dips are recognizable (i.e., identifiable) for weeks and possibly as long as several months. These findings indicate a certain stability to the dust clouds which strongly suggests that the dust cannot be formed at a single instant, with the resultant dust cloud surviving as a discrete feature for many orbits.  The argument against such a scenario is the following.  Consider a dust cloud that forms and grows to have a characteristic size that is comparable to that of the white dwarf.  In that case, the cloud size likely extends in the radial direction (i.e., toward and away from the WD) by a distance that is of the order of 1\% its orbital radius.  In turn, this means that the spread in orbital frequencies is $\sim$1.5\% of the mean orbital frequency.  In that case, the dust cloud would spread out along its orbit by 360$^\circ$ in about 65 orbits.  However, most of the observed dips have a duration much less than one hour, which is only $\sim$20\% of an orbital cycle.  This would occur within $\sim$ a dozen orbits, or within a couple of days.

Since we are able to track many of the dip features for weeks, to even months in several cases, this implies that the dust must be at least partly continually produced.  Whether this dust production is due to sublimation from the molten surfaces of asteroids, or a nearly continuous bombardment of an asteroid by smaller bodies, including micrometeorites, is a matter that will not be addressed in this work.

\subsection{The Case for Collisions}
\label{sec:collisions}

Collisions between solid bodies can lead to the sudden creation of a dust cloud for two reasons. (1) The collision may produce debris with a particle-size distribution that extends down to micron-size dust. (2) Fresh surfaces are exposed to irradiation from the WD, which leads to further heating, melting, and sublimation. The sudden appearance of a dip, or the sudden increase in an existing dip's depth, are candidates for the collision interpretation. 

The 2016 April 20 group of dips, G6420 (see Fig.~\ref{fig:waterfall3a}), is most easily explained by a fragment/fragment collision. The dips appeared without the presence of dips at their phase location on earlier dates, so the dust clouds associated with each fragment were new. Each of the four major fragments was observed to orbit with a different period, and this indicates that they gained or lost specific orbital angular momentum and/or energy due to the collision. The case for a fragment/fragment collision is based on the location of the dips far from the `A' asteroid's orbital position. 

The 25-fold increase in dip activity that occurred sometime between 2015 June and October is another good candidate for the collision explanation. We suggest that a fragment may have undergone a change in its orbital eccentricity due to a nearby passage with another fragment, or even the `A' asteroid (see Gurri et al.~2016), and the new orbit then had an apastron distance that was close to the asteroid's orbit. In such a scenario it would be possible for the fragment to collide with the `A' asteroid, and this could initiate the release of new fragments from the asteroid.

If this explanation is correct then we would expect to see a diverging pattern of dip drift lines, analogous to what was seen for the G6420 group. Along this same line, as shown by Fig.~6 in the R16 paper, a waterfall plot using the `A'-fragment ephemeris, 14 drift lines do in fact diverge from a date preceding the 2015 November start of observations (they project backward to a date region near 2015 August and September). 

\subsection{Dust Clouds and Debris Disk}
\label{sec:debris}

The spectral energy distribution (SED) described in Vanderburg et al.~(2015) requires a debris disk to account for the near-IR flux measurements (WISE Ch1 and Ch2, at 3.4 and 4.6 micron wavelength). A best fit was achieved for a simple uniform disk with a temperature of $1150 \pm 200$ K and a projected optically thick area $115^{+130}_{-60}$ times that of the WD (i.e., $\pi R_{\rm wd}^2$; see also Vanderburg et al.~2015). What is the connection between the particles in the debris disk and the dust clouds that produce observable dips? Even observing 870 dips, as we have in this work, will not answer this interesting question. 

From the orbital period of the dust clouds, combined with the estimated WD mass, we know that the dust clouds orbit at a distance from the WD of $\sim$94 WD radii ($94 \, R_{\rm wd}$). If the `A'-fragments are coplanar with the debris disk (which assumes the debris is disk-shaped), we might surmise that the `A'-fragments define the inner edge of the debris disk. The asteroids presumed to exist corresponding to the K2 periods `B' through `F' would then extend out to orbital radii of 100 $R_{\rm wd}$, which would presumably still lie well within the debris disk.  In order for bodies orbiting at 100 $R_{\rm wd}$ to transit the WD, their orbital inclination angle would have to be $\gtrsim 89.4^\circ$.  If an optically thick dust disk, responsible for the near-IR excess, is coplanar with these orbiting asteroids, and if the inner edge of the dust disk is at 94 $R_{\rm wd}$, then the outer edge of the disk would lie at $\sim$$140^{+40}_{-20} \,R_{\rm wd}$.  This outer edge of the disk is set simply to achieve the requisite projected area of $\sim$115 $\pi R_{\rm wd}^2$ at an inclination of 89.4$^\circ$. 

If the asteroids, and their fragments (which are presumed to be the source of the dust clouds), have an inclination of $89.4^\circ$, the dust clouds would have to extend to at least $\sim$1.2 $R_{\rm wd}$ from the orbital plane in order to produce the transit depths we observe.  For inclination angles as low as $89^\circ$ or $88^\circ$ this would require dust clouds extending 2.9 $R_{\rm wd}$ and  4.6 $R_{\rm wd}$, respectively, from the orbital plane.  In turn, the projected area of such dust clouds would then be $\gtrsim$ 9  and $\gtrsim$ 20 times that of the white dwarf, if the dust clouds are at least as wide as they are high. Our data cannot rule out such a geometry; however, the implications for dust production rates are greater, and it will be for future modeling to assess whether the geometry associated with inclinations as low as $88^\circ$ are feasible.  

\subsection{Fraction of WDs Like WD1145}

The K2 observations included $\sim$860 WDs whose flux light curves were analyzed in a manner similar to that for WD1145, but the WD1145 dips remain unique (A. Vanderburg, 2016, and private communication 2016). The simplest interpretation is that if $\mathcal{F}$ is the fraction of WDs meeting the dual conditions of (i) being orbited by dust clouds and (ii) having inclinations close enough to 90$^\circ$ for transits,
%any dust clouds that are present to be observed as transit dips, 
then $\mathcal{F} \simeq 1/860 \simeq 0.12\%$ with a 90\% confidence range of $0.04\% - 0.55\%$. If 1\% of inclinations are acceptable for producing transits when dust clouds are present, then perhaps $\sim$4\%-55\% of WDs (i.e., $\mathcal{F}/0.01$) may have orbiting dust clouds. This statistic is compatible with the finding that 25 to 50\% of WDs have spectra exhibiting atmospheric pollution by metal absorption lines. We do not know what fraction of the time such stars exhibit dramatic increases in activity, such as the 25-fold one that we observed for WD1145 as Event ``1'' (Fig.~\ref{fig:activity2}), but if it is 10\%, for example, then among the 13,000 known WDs\footnote{\url{http://www.astronomy.villanova.edu/WDCatalog/}} there should be $\sim$$1-7$  WDs (i.e., $1.3\times10^4 \times 0.1 \times \mathcal{F}$) currently undergoing high levels of dust-produced transit activity.  Future surveys will be able to provide a more accurate estimate of how often WDs with dust clouds exhibit the kind of easily observed, high levels of  activity reported in this paper for an 8-month observing interval.

\subsection{Future Role for Amateur Observations}

Among the four surprise events mentioned above, three were identified and well documented because the lightcurve observations of this work were frequent, sometimes with a daily cadence (see, e.g., Figs.~\ref{fig:waterfall} and \ref{fig:activity1}). Such monitoring is feasible for non-professional observers, provided dip depths exceeding $\sim$3\% are present and are sufficient for characterizing dip activity. This was the situation beginning after the 2015 October announcement that WD1145 was being transited by dust clouds, and it was superb good luck that when amateur observations began (2015 November 01) the dip activity level was suitably high for useful amateur observation and frequent monitoring, and remained at a heightened level for most of the remaining observing season. If the heightened activity had occurred a year earlier only professional telescopes would have been in use for photometric observing, and the frequency of lightcurve observations might have been inadequate for the quality of characterization that is reported here.  

If the trend of decreasing activity since 2016 April continues, by the start of the next observing season (2016 November) WD1145 will have completely returned to the {\em Kepler} K2 observing-epoch activity level. WD1145 has the prospect of becoming a regular target for advanced amateurs, and for pro/am collaborations. Part of the appeal for dedicated observation is that the lightcurves change very often and dramatic surprises can be expected to occur several times a year during the heightened activity state. Continued amateur monitoring can be counted on to provide news of the next outburst of heightened activity.

\section{Summary and Conclusions}
\label{sec:concl}

In this work we have documented the photometric behavior of WD1145 during an 8-month period of high dip activity.  The total data set comprises 158 lightcurves collected over the 8-month interval during which we detected $\sim$800 significant dips in flux.  We present an illustrative set of the lightcurves which show how the dips vary from night to night and from month to month.  We use these to construct `waterfall diagrams', which in turn allow us to track specific repeatable dip features in the lightcurves over substantial periods of time (days to months).  The features that can be tracked lead to precise orbital periods of the bodies responsible for producing the dust, which we infer to be fragments of asteroids.

During the 8-month interval, we observed several very notable changes in the dip behavior. First, the earliest amateur observations, in 2015 November, showed that WD1145 was producing many more dips per orbit than for all previous observations. Second, in 2016 January a large number of dust clouds appeared that had an orbital period of approximately 4.4912 hours. This event marked the end of a 3-month interval of short-lived dips that produced an apparent 4.5004-hour periodicity which previous studies had associated with the {\em Kepler} K2 `A' period. The third noteworthy event was a 2016 April 21 appearance of four dip features with drift lines in a waterfall diagram that diverged from their origin date and lasted for about two weeks.  These dips appeared at a location in the orbit that cannot be explained as fragments that had just broken away from the `A' asteroid. The fourth event was the sudden appearance of a dip feature with a period of 4.6064 hours, which we associate with the Kepler K2 `B' period. This period had not been detected since the original K2 observations.  Perhaps a final ``surprise'' is the apparent decline of dip activity during 2016 May, June and July to a level that is almost as low as that observed by {\em Kepler}, two years earlier, and by professional astronomers in early 2015.

We made use of the quantitative dip properties to define an ``equivalent width" (`EW') for all the dips in a given orbit -- basically the integrated area under the normalized flux lightcurve.  In this way, we could quantitatively follow the source `activity' for all the dips during an orbit as a function of date.  We found that during most of our observations the overall dip activity level was more than an order of magnitude higher than during the time it was discovered with K2. 

We employed several different algorithms to search for periodicities in the entire 8-month data set.  These include searches by eye for related dip-drift features, the BLS algorithm, a search for inter-dip times that match trial periods, and a Hough transform.  The one consistent period that appears in all the searches is 4.4912 hours.  We associate this largely with the G6121 set of dips.  We also detect a few other periodicities which may not be as coherent as that associated with the G6121 dips; these are displayed graphically in Fig.~\ref{fig:Pfrag}. Finally in regard to periodicities, we believe we have detected the K2 `B' period.

Given the exciting developments following the discovery of WD1145, the notion that WDs with ``polluted atmospheres'' (i.e., exhibiting metal-line absorption spectra) are orbited by asteroid-size bodies in a way resembling WD1145, becomes ever more compelling. It is also quite reasonable to assume that approximately 1\% of such systems are oriented so that an asteroid and its fragments transit the WD (orbital inclination $\gtrsim$ 89 degrees).  However, it is not known how often such WDs exhibit the same high level of dip activity that was encountered during the 8 months of observations described in this work. The presence of dips with 1 or 2\% depth, as measured by {\em Kepler,} may be the most common state.  Therefore, as we argued, it is plausible that of the $\sim$13,000 known WDs, there may indeed be a handful that are currently in the dipping state of WD1145. 

Data files for the 6 months of observations of this work, plus the 2.5 months of observations reported on in the R16 publication, are available upon request from author BLG. These files include normalized flux for each lightcurve data point, for each of the 158 lightcurves, as well as their 870 dip solutions (BJD, depth, $\tau_1,  \tau_2$, observer and telescope for each dip). Web page URLs are also available from BLG that show all lightcurves in great detail and in several formats. 

\vspace{0.3cm}
\noindent
{\bf Acknowledgements}

We thank Andrew Vanderburg and Enric Palle for helpful discussions. The amateur observers (B.\,L.\,G., T.\,G.\,K., F.-J.\,H.) are self-funded so for them no institutional or tax-funded agencies need acknowledgement. The IAC80 is operated on the island of Tenerife by the Instituto de Astrof\'\i sica de Canarias in the Spanish Observatorio del Teide. RA acknowledges the work of the ``T\'ecnicos en operationes telesc\'opicas'' and the rest of the OT staff for continuous operation of the facilities. RA acknowledges the Spanish Ministry of Economy and Competitiveness (MINECO) for the financial support under the Ramon y Cajal program RYC-2010-06519, and the program RETOS ESP2014-57495-C2-1-R. We thank Paulo Miles for providing data for two nights at the IAC80. Two lightcurves were contributed by Yenal Ogmen, with a 14$''$ telescope in Cyprus, and one lightcurve was submitted by Paul Benni, using a 14$''$ telescope in Acton, MA, USA, for which we are grateful. TGK thanks Beth Christie and Cheryl Healy for material support in JBO telescope operations and equipment.

B. L.~Gary \thanks{ Email: BLGary@umich.edu, }
S.~Rappaport \thanks{ Email: sar@mit.edu, }
T.G.~Kaye \thanks{ Email: tom@TomKaye.com, }
R.~Alonso \thanks{ Email: ras@iac.es, }
and F.-J.~Hambsch \thanks{ Email: hambsch@telenet.be}

%%%%%%%%%%%%%%%%%%%%%%%%%%%%%%%%%%%%%%%%%%%%%%%%%%



%\bibliographystyle{mnras}


%%%%%%%%%%%%%%%%%%%%%%%%%%%%%%%%%%%%%%%%%%%%%%%%%%

%%%%%%%%%%%%%%%%% APPENDICES %%%%%%%%%%%%%%%%%%%%%

%\appendix

%\section{Any Appendices Go Here}
%\label{app:1}

%%%%%%%%%%%%%%%%%%%%%%%%%%%%%%%%%%%%%%%

% Don't change these lines
\bsp	% typesetting comment
\label{lastpage}
\end{document}